\documentclass{aastex62}

\usepackage{booktabs}

\usepackage{lmodern}
\usepackage{textcomp}
\usepackage[T1]{fontenc}
\usepackage{amsmath}
\usepackage[english]{babel}
\usepackage{float}
\usepackage{hyperref}
\usepackage{amssymb}
\usepackage{aas_macros}
\usepackage{relsize}
\usepackage[innercaption]{sidecap}
\usepackage{array}   
\newcolumntype{L}{>{$}l<{$}}
\newcolumntype{C}{>{$}c<{$}}

\newcommand{\sd}{\texttt{shaddisk} }
\newcommand{\xspec}{\textsc{Xspec} }

\newcommand{\rns}{$ R_{NS} $}

\newcommand{\kalp}{K$\alpha$}

\newcommand{\sedi}{4U 1636-53}

\graphicspath{{./}{figures/}}

\received{in 2019; this preprint edition is presented as it was submitted, \textit{i.e.} before the small additions suggested by the referee}

\submitjournal{ApJ}

\shorttitle{Neutron star R/M from partially occulted Fe-line profiles}
\shortauthors{La Placa et al.}

\begin{document}
	
\title{Neutron Star Radius-to-mass Ratio from Partial Accretion Disc Occultation as Measured through Fe K$\alpha$ Line Profiles}

\correspondingauthor{Riccardo {La Placa}}
\email{riccardo.laplaca@inaf.it}

\author{Riccardo {La Placa}}
\affil{Department of Physics G. Marconi,
	University of Rome ``La Sapienza'', Italy}
\affiliation{INAF - Astronomical Observatory of Rome,
	Monte Porzio Catone (Roma), Italy}
\affiliation{Research Centre for Computational Physics and Data Processing, Silesian University in Opava, Czech Republic}

\author{Luigi Stella}
\affiliation{INAF - Astronomical Observatory of Rome,
	Monte Porzio Catone (Roma), Italy}

\author{Alessandro Papitto}
\affiliation{INAF - Astronomical Observatory of Rome,
	Monte Porzio Catone (Roma), Italy}

\author{Pavel Bakala}
\affiliation{Research Centre for Computational Physics and Data Processing, Silesian University in Opava, Czech Republic}
\affiliation{M. R. \v{S}tef\'anik Observatory and Planetarium, Hlohovec, Slovak Republic}

\author{Tiziana {Di Salvo}}
\affiliation{Universit\`{a} degli Studi di Palermo, Dipartimento di Fisica e Chimica, Palermo, Italy}

\author{Maurizio Falanga}
\affiliation{International Space Science Institute (ISSI), Bern, Switzerland}
\affiliation{International Space Science Institute Beijing, PR China}

\author{Vittorio {De Falco}}
\affiliation{Research Centre for Computational Physics and Data Processing, Silesian University in Opava, Czech Republic}

\author{Alessandra {De Rosa}}
\affiliation{INAF/Istituto di Astrofisica e Planetologia Spaziali, Roma, Italy}

\begin{abstract}

	We present a new method to measure the radius-to-mass ratio (R/M) of weakly magnetic, disc-accreting neutron stars by exploiting the occultation of parts of the inner disc by the star itself. This occultation imprints characteristic features on the X-ray line profile that are unique and are expected to be present in low mass X-ray binary systems seen under inclinations higher than \textasciitilde$65$~degrees. We analyse a \textit{NuSTAR} observation of a good candidate system, \sedi, and find that X-ray spectra from current instrumentation are unlikely to single out the occultation features owing to insufficient signal-to-noise. Based on an extensive set of simulations we show that large-area X-ray detectors of the future generation could measure R/M to \textasciitilde$2\div3$\% precision over a range of inclinations. Such is the precision in radius determination required to derive tight constraints on the equation of state of ultradense matter and it represents the goal that other methods too aim to achieve in the future.
\end{abstract}


\section{Introduction} \label{sec:intro}
The central regions of neutron stars (NSs) attain densities close to or exceeding those of atomic nuclei. They provide unique laboratories for investigating the physics of dense and cold bulk nuclear matter, in which the strong force contributes to the pressure and exotic states might be present \citep[for a review see, e.g.,][]{LattPrak2016, Watts2016}. 
A key diagnostic of dense matter is its equation of state (EoS), i.e. the pressure-density relation. A variety of EoS have been proposed, each representing a different possible theory describing the properties of dense matter, which are still consistent with data from heavy-ion colliders and other experiments \citep[see, e.g.,][]{LattPrak2016, ASSL2018}. Based on them, the Tolman-Oppenheimer-Volkoff equations are integrated and present-day neutron star models and sequences are calculated for each EoS. The predicted macroscopic properties of NSs can be tested against astrophysical measurements and constraints. In particular if some key observables, chiefly neutron star masses, M, and radii, R, can be accurately measured, it is then possible to constrain the EoS. In fact different EoS predict different M-R relations: for example models based on GS1 \citep{Glendenning1999} contain large amounts of exotic particles which ``soften'' the EoS, giving rise to relatively small radii and maximum masses; on the other hand, strange quark matter (SQM) stars with low masses are almost incompressible \citep{LattPrak2004}. A variety of nucleonic neutron star models predict the radius to be nearly insensitive to the mass around \textasciitilde$ 1\div1.5~M_{\odot}$ \citep[e.g. MPA1, MS2, AP3, respectively][]{Muther1987,Muller1996,Akmal1998}, which makes the determination of the radius of stars in that range a crucial factor in discerning among different EoS.

Two different approaches have been discussed in EoS testing. The direct approach consists in determining the likelihood that the measured M and R pairs match the M-R relation predicted by a given EoS. The other approach, pioneered by \citet{Lindblom1992}, involves the inverse process, namely mapping the measured M and R pairs to the EoS. Based on a parametrised representation of the EoS, \citet{Ozel2009} showed that several \textasciitilde$5\%$ precise measurements of M and R pairs are required to discriminate between EoS models with 3$\sigma$ confidence.  

Neutron star masses have been measured for a number of radio pulsars and X-ray binary systems \citep[for a review see][]{Ozel2016} and, more recently, also in the merging neutron star binary GW170817 \citep{Abbott2018Mass}. The highest precision mass measurements are those from a few radio pulsar binaries containing two neutron stars \citep[see, e.g.,][]{Burgay2003, Lyne2004}.

Radii are far more difficult to measure; different techniques, mostly based on X-ray diagnostics, have been devised to determine either the neutron star radius itself or its ratio to the mass.  
X-ray spectroscopy-based techniques include (a) redshifted ion lines from the neutron star surface \citep[if any, e.g.][]{Cottam2002}, (b) the flux and temperature of thermal emission from quiescent neutron star transients \citep[e.g.][]{Heinke2014} and (c) the evolution and ``touch down'' phase of thermonuclear flashes that give rise to photospheric radius expansion in accreting neutron stars \citep{Lewin1993}; the latter method provides simultaneous mass and radius measurements.
Other possible techniques are based on timing diagnostics \citep[see, e.g.,][]{Bhattacharyya2010,Watts2016}. 
For instance, (d) the fastest neutron star spin periods \citep[e.g.][]{Haensel2009}, as well as (e) the fastest quasi-periodic oscillation signals observed in X-ray binaries \citep[if arising from Keplerian motion: see, e.g.,][\S 3.3.2 and references therein]{Bhattacharyya2010} provide mass-dependent upper limits on the radius which allow to exclude entire regions of the M-R diagram; (f) the quasi-periodic X-ray signals observed during giant flares of magnetars, if asteroseismic in origin, hold the potential to measure both M and R \citep{Israel2005,Steiner2009}; (g) a promising timing technique for simultaneous measurements of M and R exploits the X-ray modulation generated by hotspots on the surface of isolated or accreting neutron stars spinning with periods in the millisecond range \citep[e.g.][]{Nattila2018}. 

Along with these techniques, entirely different ones have been proposed to constrain the EoS of 
ultradense matter in neutron stars: (h) the measurement of a star's moment of inertia in relativistic radio pulsar binaries hosting two neutron stars, especially PSR J0737-3039A/B \citep{LattSchutz2005, Kehl2016}; (i) the study of the tidal deformability in the merging events of two neutron stars through their gravitational wave signal \citep{Hinderer2010, Raithel2018}; (j) the analysis of the peak frequency of the post-merging gravitational wave signal, if an hypermassive, differentially rotating neutron star is formed in the merging \citep{Chatziioannou2017}. 

Despite considerable progress in recent years, radius (or combined mass and radius) measurements have not yet attained the required level of accuracy and precision to univocally determine the EoS \citep[e.g.][]{Watts2019}. Limitations involve systematics, insufficient signal-to-noise or resolution, modelling uncertainties and scarcity of suitable systems or events.

In this paper we introduce a new X-ray spectroscopic technique aimed at determining the radius-to-mass ratio of neutron stars accreting through a disc in low mass X-ray binaries (LMXBs). 
The technique exploits the very broad and redshifted profile of the Fe \kalp\ line around 6~keV that is observed from a number of neutron star and stellar mass black hole LMXBs, as well as AGNs. There exists by now a large body of evidence that these lines originate in the innermost disc regions, where their profile is determined by relativistic beaming, time dilation, red/blue-shifts, light bending and frame dragging of matter orbiting the inner regions of accretion discs at non-negligible fractions of the speed of light, $c$ \citep[][]{Fabian1989,Reynolds2003,Reynolds2014}. The inner radius of the accretion disc is one of the key parameters that is routinely derived from combined fits to the Fe \kalp\ profile and X-ray spectral continuum; in application to accreting neutron star systems inner disc radii as low as \textasciitilde$ 6 \div 7\ GM/c^2 $ were obtained in some cases, which provide upper limits on the star's radius-to-mass ratio \citep[see, e.g.,][]{Bhattacharyya2011,Ludlam2017}.

Our new technique applies to LMXBs seen under high inclinations ($i \gtrsim 65$\textdegree) in which the line-emitting disc extends to very close to the neutron star surface; in those systems we expect the body of the neutron star to occult the line flux from that part of disc innermost region that is behind the NS, giving rise to distinctive features in the Fe \kalp\ line profile which encode precise information on the R/M ratio. 

Our paper is structured as follows: in Section 2 we describe our model and its approximations; in Section 3 we discuss the effects of occultation on different line profiles; Section 4 presents the first application of our technique to fit the X-ray spectrum of a NS LMXB system, 4U1636-53, as determined from an observation with the \textit{NuSTAR} satellite; we also investigate the precision of the radius-to-mass ratio measurements that can be obtained with large area X-ray instruments of the next generation, in light of the extensive simulations we carried out; in Section 5 we discuss the limitations of our technique and outline future perspectives.

\section{Line Profile Calculation} \label{sec:line}

Our technique adopts a general relativistic approach in the Schwarzschild metric, exploits a high-precision approximation of strong-field photon deflection in the Schwarzschild spacetime and accounts for the angular dependence of the disc emissivity. The line flux measured by an observer at infinity at a given frequency $\nu_{obs}$ is \citep{Misner1973}
\begin{equation}\label{eq:flux}
F_{obs}(\nu_{obs}) = {\int } \dfrac{I_{\nu_{em}}}{(1+z)^{3}} d\Omega,
\end{equation}
where $d\Omega $ is the solid angle subtended by the disc element in the observer's sky, 
$ I_{\nu_{em}} $ is the disc specific emissivity at the emission frequency, $(1+z)^{-1}=\nu_{obs}/\nu_{em} $ is the redshift factor and the integration extends over the whole disc surface contributing to the flux at $\nu_{obs}$.

\begin{figure}[t!]
	\centering
		\gridline{\hspace{-0.03\textwidth}
			\fig{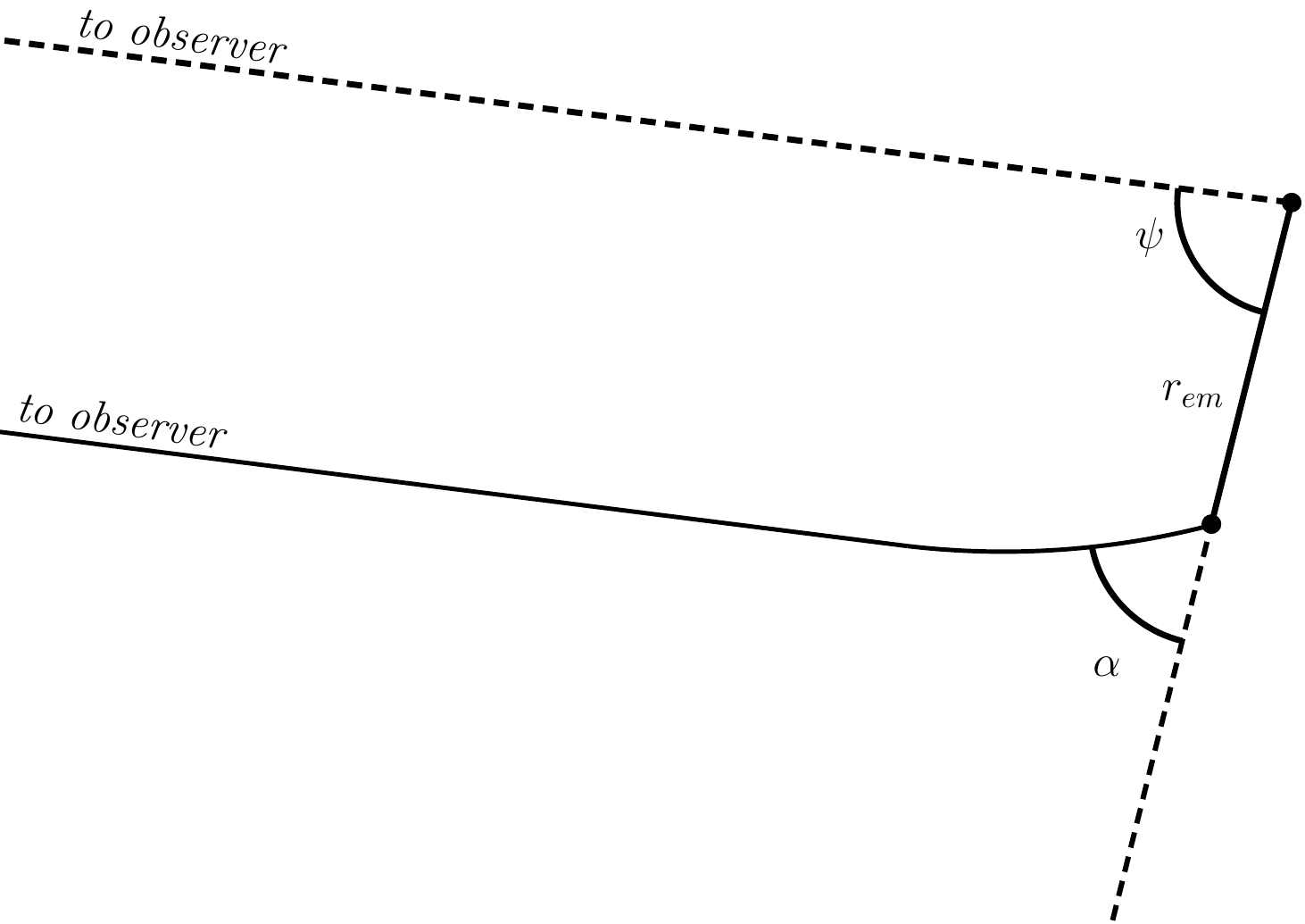}{0.51\textwidth}{}\hspace{0.04\textwidth}
			\fig{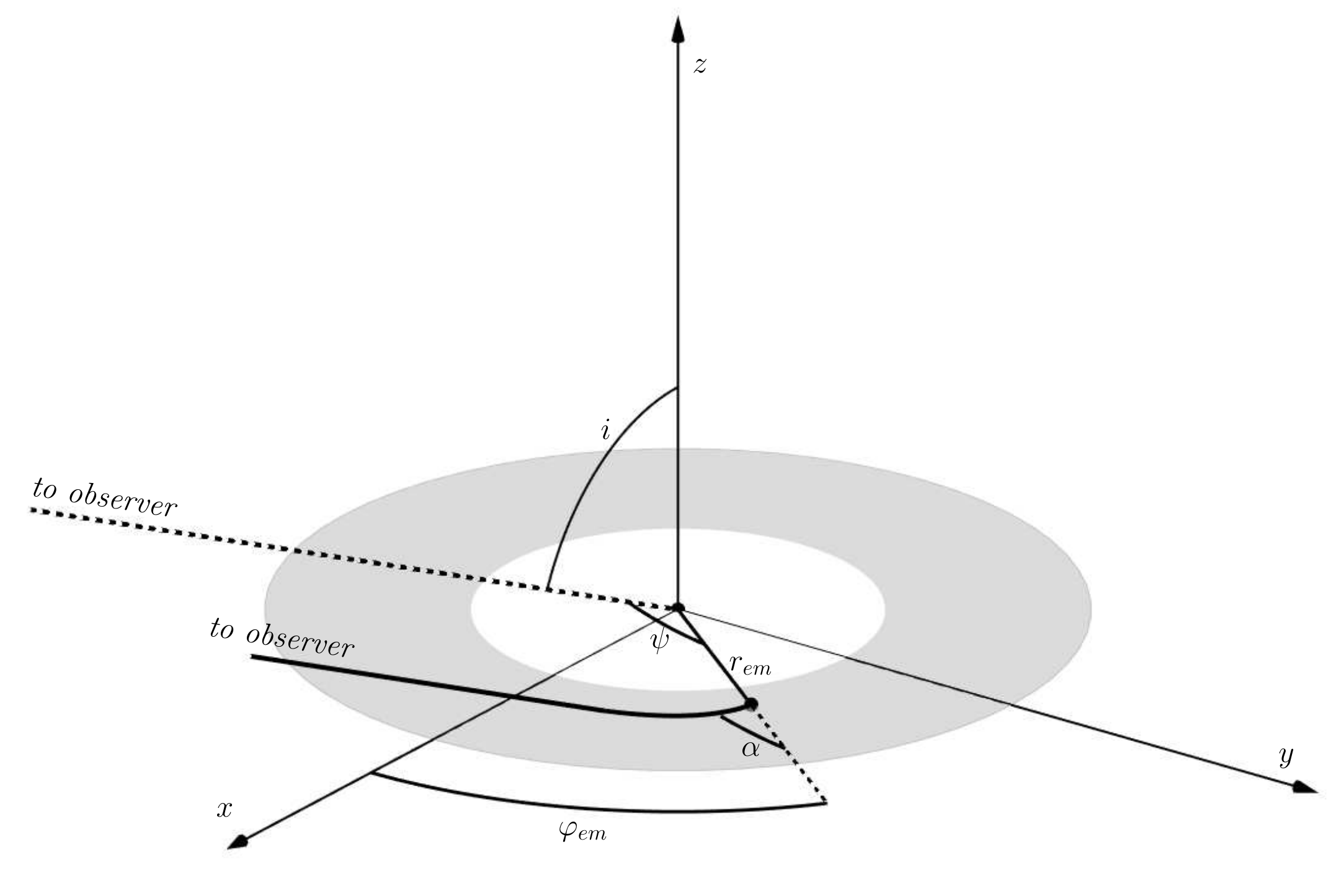}{0.51\textwidth}{}
		}
	\caption{We show here a sample photon trajectory plane (left) and the geometry of our system (right): the observer, and thus the line of sight, lies on the $ xz $ plane while angles $ \alpha $ and $ \psi $ lie on the photon trajectory plane, which also contains the line of sight. The inclination is denoted as $ i $ and the coordinates of the emitting point on the disc as $ (r_{em}, \varphi_{em})$.}
	\label{fig:geom1}
\end{figure}

Equation \ref{eq:flux} also includes information on the photon trajectories connecting emission points on the disc, here assumed to lie in the equatorial plane and have negligible vertical thickness ($ \theta_{em} = \pi/2 $ for any emission point), to the observer. 
Photons emitted at $r_{em}$ and $\varphi_{em}$ travel along null geodesics lying on the plane defined by the lines connecting the origin to the emitting point and to the observer at infinity (the line of sight). We characterise these geodesics through the angle between the two aforementioned lines, $ \psi $, and the angle between the emission direction and the radial direction, $ \alpha $ (see Fig. \ref{fig:geom1}). Photon geodesics starting at $r_{em}$ are described by 
\citep{Chandrasekhar1983, Beloborodov2002}
\begin{equation}\label{eq:lightbending} \psi = {\int }_{r_{em} }^{\infty } \left[\frac{1}{b^{2}} - \frac{1}{r^{2}} \left(1-\frac{R_{S}}{r}\right)\right]^{-\frac{1}{2}}\frac{dr}{r^{2}} , \end{equation}
where $ b $ is the impact parameter of the photon trajectory and $ R_{S} = 2GM/c^2$ represents the Schwarzschild radius. Throughout this paper we use geometric units ($G=c=1$) and express lengths in units of the gravitational radius $ GM/c^2 = r_{g} = R_{S}/2 $. The impact parameter is related to $ \alpha $ by 
\begin{equation}\label{eq:imppar} b = \frac{r_{em} \sin \alpha}{\sqrt{1 - R_{S}/r_{em}}} \end{equation}
and from spherical trigonometry we obtain
\begin{equation}\label{eq:geompsi}
\cos \psi = \sin i   \sin \theta   \cos \varphi   +   \cos i   \cos \theta.
\end{equation} 

A distinction must be made between photons with direct trajectories ($0 < \alpha < \pi/2$) and photons which go through a turning point ($\alpha > \pi/2$): the former always reach the observer, whereas those among the latter that reach infinity have impact parameters $b > b_{min} = 3\sqrt{3}~r_{g}$ \citep{Misner1973}. This last condition translates into a limit on the maximum $ \alpha $ angle through Eq. \ref{eq:imppar}; in subsection \ref{ssec:occ} we discuss the way in which the above limit changes in the presence of a star whose radius exceeds $3~r_{g}$.

Assuming that the disc matter moves along circular Keplerian orbits, the redshift factor of photons by any point on it is given by \citep[e.g.][]{DFS2016}
\begin{equation}\label{eq:redshift}
(1+z)^{-1} = \dfrac{\sqrt{1 - 3r_{g}/r_{em}}}{1 + \dfrac{b}{r_{em}} \sqrt{\dfrac{r_{g}}{r_{em}}}\dfrac{\sin(i) \sin(\varphi_{em})}{\sin(\psi)}} \ .
\end{equation} 

The solid angle element in the observer's sky is given by \citep{Bao1994, DFS2016}
\begin{equation}\label{eq:solid1}
d\Omega = \dfrac{b~db~d\varphi'}{D^{2}} = \dfrac{b}{D^{2}} \dfrac{\partial \varphi'}{\partial \varphi} \dfrac{\partial b}{\partial r} dr d\varphi
\end{equation}
where $\varphi'$ is calculated in the observer's frame, $ D $ is the distance to the system and the Jacobian is calculated through the partial derivatives $ \partial b / \partial \psi $ and $\partial \psi / \partial r$ by means of Eq. \ref{eq:lightbending}; its explicit form is 
\begin{equation}\label{eq:solid2}
d\Omega = \dfrac{\dfrac{b^{2} \cos i}{r_{em}^{2} D^{2} \sin^{2}\psi \ \cos \alpha}}{\mathlarger{\int }_{r_{em}}^{\infty } \left[1 - \dfrac{b^{2}}{r^{2}} \left(1-\dfrac{R_{S}}{r}\right)\right]^{-\frac{3}{2}}\dfrac{dr}{r^{2}}} dr d\varphi .
\end{equation}

The specific intensity emitted in the local corotating frame of the disc is written as
\begin{equation}\label{eq:emissivity}
I_{\nu_{em}} = \delta(\nu - \nu_{em})\mathcal{R}(r_{em}) \mathcal{A}(\cos \lambda_{em}),
\end{equation}
where $ \nu_{em} $ is the laboratory frequency of the emitted photon.

The radial emissivity law, $\mathcal{R}(r_{em})$ is determined by the source of illumination of the disc and is usually approximated by a power law $ \mathcal{R}(r_{em}) \propto (r_{em})^{q} $, with $q$ in the $ (-3 \div -2) $ range. In the so-called \textit{lamp-post} geometry, a point-like source placed along the z-axis above the compact object \citep[see, e.g.,][]{Martocchia1996}, we have $ q = -3 $ for $r_{em}\geqslant 10~r_{g}$ and a somewhat flatter dependence at smaller radii. The radial emissivity can be approximated with a single $q\simeq -3$ power law (or a broken power law) also when the illuminating source is not point-like but extended, as in the case of a hot gas corona hovering above the inner parts of the disc or enshrouding the neutron star \citep{Wilkins2018}. An illuminating equatorial bright belt on the neutron star surface, resulting from energy released in the boundary layer between the accretion disc and the star, would present instead a marked steepening in the emissivity law at the smallest radii  $r_{em}\leqslant 8~r_{g}$ \citep[see fig. 5 in][]{Wilkins2018}.

The angular dependence $ \mathcal{A}(\mu = \cos \lambda_{em}) $ of the local disc emissivity (here $ \lambda_{em} $ is the angle between the photon emission direction and the disc normal, as measured in the corotating reference frame) is determined by the physical processes governing the production of line photons. We consider three different cases: limb-darkened, isotropic and limb-brightened emission, the latter being favoured when fluorescent emission results from illumination by an external X-ray source.

The limb-darkening law is modelled after \citet{Chandrasekhar1950} assuming a pure scattering atmosphere, and its most commonly used form is $\mathcal{A}(\mu) \propto 1 + 2.06\mu$ \citep{Laor1991}: the conditions that would produce such an emissivity, however, are difficult to warrant at all disc radii \citep[see, e.g.,][and references therein]{SvobodaPhD}. Later works showed that a limb-brightening law would be more apt \citep{Haardt1993, Goosmann2007}. This case is represented by an angular dependence $ \mathcal{A}(\mu) \propto \log(1 + 1/\mu)$ \citep{Haardt1993}, and it is also consistent with the results of Monte Carlo simulations of accreting systems \citep{George1991, MattPP1991, Ghisellini1994}. Lastly, the law for isotropic emission $ \mathcal{A}(\mu) =$ \textit{const.} is more appropriate for low optical depths and no internal heating of the disc atmosphere \citep{Fukue2006}, and can provide an intermediate case if the system configuration is poorly known, especially because the degree of ionisation of the disc components is expected to play a role in the observed angular emissivity \citep[see, e.g.,][]{Goosmann2007}.

In Schwarzschild spacetime $ \mu $ is given by \citep[see, e.g.,][]{ChenEardley1991, Bao1994} 
\begin{equation}\label{eq:angdep} \mu = \dfrac{b/r_{em}}{(1+z)} \dfrac{\cos(i)}{\sqrt{1 - \sin^2 (i) \cos^2 (\varphi_{em})}}  
= \dfrac{\cos(i) \sqrt{1-3r_{g}/r_{em}}}{\dfrac{\sin(\psi)}{\sin(\alpha)} \sqrt{1-\dfrac{2r_{g}}{r_{em}}} +\dfrac{\sin(i) \sin(\varphi_{em})}{\sqrt{r_{em}/r_{g}}}} \ : \end{equation}
as expected, $ \mu $ tends to $ \cos(i) $ when $ r_{em}\to\infty $.

\subsection{Analytical approximation of the photon geodesic equation}\label{sec:approx}

In order to avoid numerical integration of the photon geodesics, which would slow down the code considerably, we adopt an analytical approximation for Eq. \ref{eq:lightbending} \citep{ResNote}:
\begin{equation}\label{eq:approxgeo}
1 - \cos \alpha \approx (1 - \cos \psi)(1 - \dfrac{2r_{g}}{r})[1 + k_{1} \dfrac{2r_{g}}{r} (1 - \cos(\psi-k_{2}))^{k_{3}}],
\end{equation} where  $k_{1} = 0.1416,~k_{2} = 1.196$ and $k_{3} = 2.726$. This equation couples $ \alpha $ to $ \psi$ and, together with Eq. \ref{eq:imppar} and \ref{eq:geompsi}, it links the coordinates of an emitting point on the disc to the impact parameter.

The solid angle element (Eq. \ref{eq:solid2}) can be approximated by using Eq. \ref{eq:approxgeo} to calculate only the derivative $ \partial b/\partial r $ \citep[see, e.g.,][]{Bao1994}, since the other terms in Eq. \ref{eq:solid2} already are analytical.

\subsection{Occultation}\label{ssec:occ}

If the disc inner radius lies close to the surface of the star and the inclination under which we observe the system is high enough, some points on the far side of the disc ($x < 0$) will be occulted by the star itself notwithstanding the gravitational light bending.
\begin{figure}[t]
	\centering
	\includegraphics[width=0.8\linewidth]{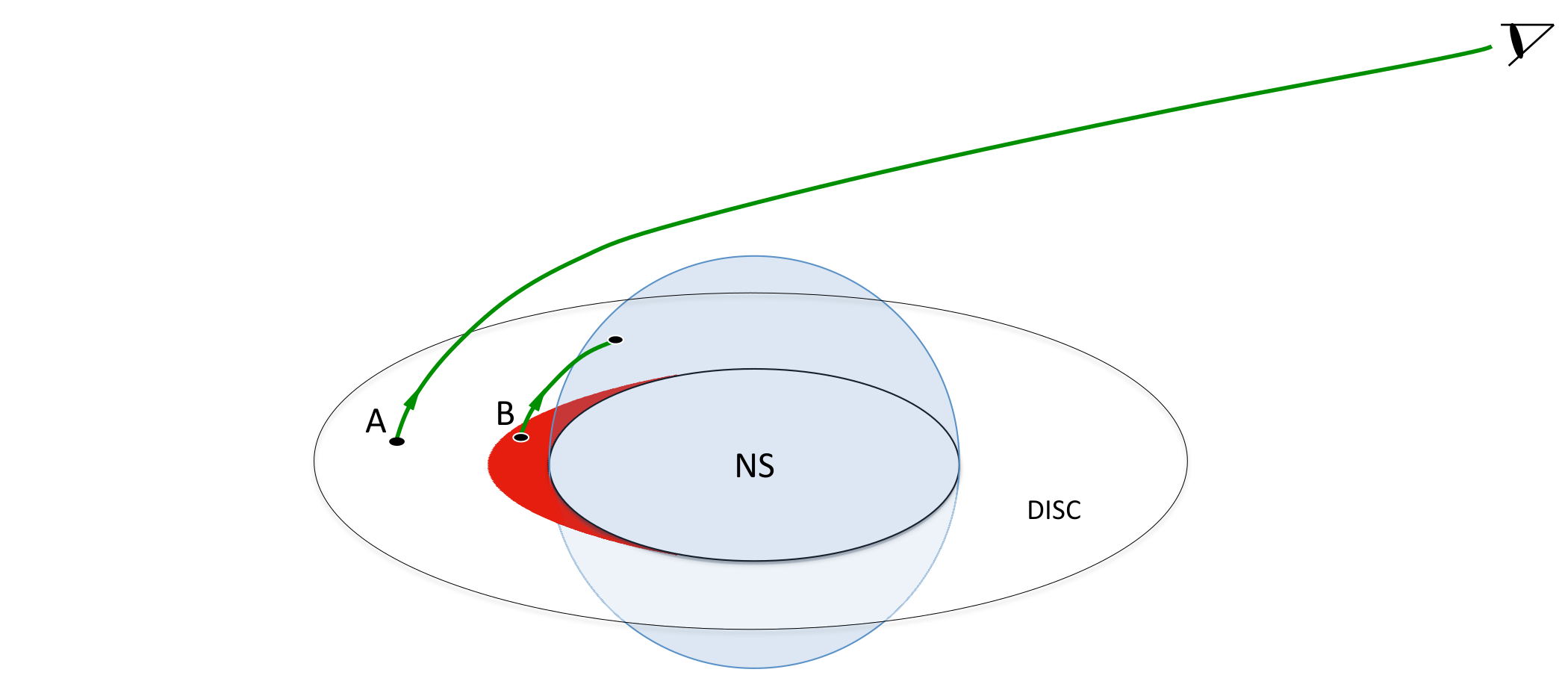}
	\caption{Simplified representation of the star body occulting part of the disc: photons starting from A reach the observer, whereas the ones emitted from the red ``crescent'' are intercepted by the star since their periastron falls beneath the star surface.}
	\label{fig:occult}
\end{figure}

\begin{figure}[t]
	\centering
	\vspace{-0.3cm}
	\includegraphics[width=0.8\linewidth]{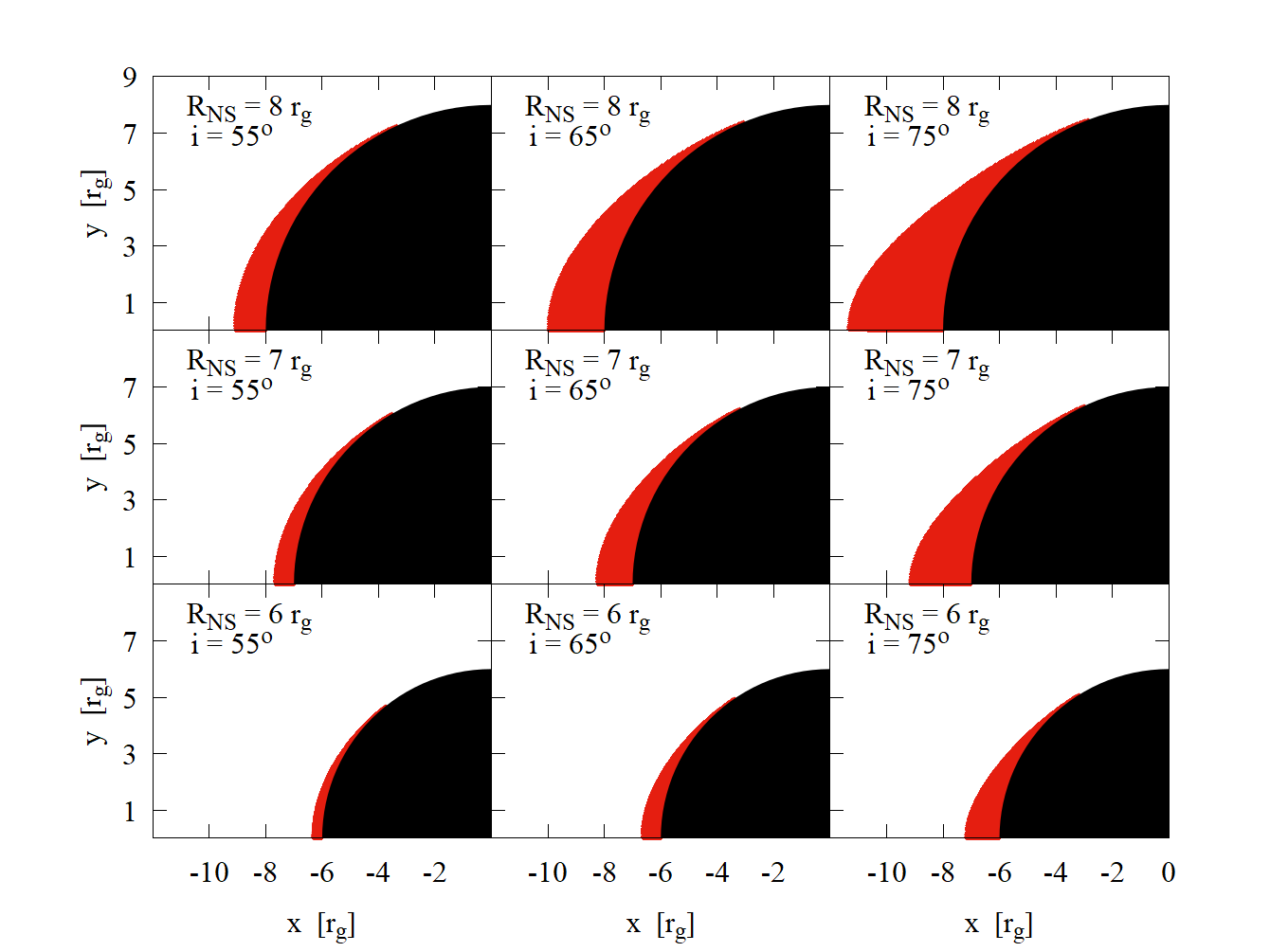}
	\caption{The nine panels show the extent of the occulted area (red) on discs that reach the neutron star (black), for different inclinations and star radii; the observers lie at infinity on the right, on the $xz$ plane, so that the quadrants show half of the ``back'' side of the star relative to them (90\textdegree\ $ \leqslant \varphi_{em} \leqslant 180$\textdegree).}
	\label{fig:shadmap}
\end{figure}

The periastron for a given impact parameter $ b $ reads 
\begin{equation}\label{eq:peri} p^{2} - (1 - \frac{R_{S}}{p})b^{2} = 0 \end{equation} and Eq. \ref{eq:imppar} describes the relation between $b$ and  $\alpha$, which in turn is connected to the coordinates on the disc through Eq. \ref{eq:approxgeo}: photon trajectories with a turning point whose periastron $ p $ falls under the star radius correspond to photons that are intercepted by the star and therefore do not reach the observer (Fig. \ref{fig:occult}). 
Plugging Eq. \ref{eq:imppar} into Eq. \ref{eq:peri} when taking $p = R_{NS}$ determines the maximum accepted value of $ \alpha $, that indicates which region of the disc should be excluded from the integration of the line profile (Fig. \ref{fig:shadmap}):
\begin{equation}\label{eq:almax} \alpha_{max}(r_{em}) = \pi - \arcsin\left[\frac{R_{NS}}{r_{em}} \sqrt{\frac{1 - R_{S}/r_{em}}{1 - R_{S}/R_{NS}}}\right]. \end{equation}

Figure \ref{fig:shadmap} shows the shape and size of the occulted disc area behind the NS for different values 
of the inclination and the star radius: as expected, the occulted area increases for increasing inclinations and star radii. 

We shall see that for high inclination and/or small inner disc radii the occultation of this crescent-like disc region by the body of the star imprints a characteristic feature on the line profile.

\section{Partially occulted line profiles} \label{sec:results}

To calculate line profiles from the model described in Sec. \ref{sec:line} we wrote a Fortran subroutine, \texttt{shaddisk}, matching the specifications of the X-ray spectral fitting program \xspec \citep{Arnaud1996}. 
\texttt{shaddisk} carries out the integration over a grid of $(r_{em}, \varphi_{em})$ on the disc. The use of Eq. \ref{eq:approxgeo} for the light bending and the solid angle element allows our code to run efficiently without resorting to numerical integration of photon geodesics or to interpolation of large matrices of pre-calculated values. 

\texttt{shaddisk}'s parameters are the rest line energy $E_o$, the radial emissivity index $ q $ (assuming a power law $ \mathcal{R}(r_{em}) \propto (r_{em} /r_{g})^{q} $), the inner and outer radii, respectively $ r_{in}/r_{g} $ and $ r_{out}/r_{g} $, the inclination $ i $ and the neutron star radius in units of the gravitational radius $ R_{NS}/r_{g} $. 
We note that this last parameter can also be thought of as the radius of an occulting sphere centred on the star, such as for instance a magnetosphere engulfed with matter of a Compton thick corona. Moreover, we inserted the possibility to choose among the three different angular emissivity laws discussed in Sec. \ref{sec:line}.

The line profiles calculated with \texttt{shaddisk} in the case of absence of occultation were tested against the profiles from \texttt{kyrline} \citep{Dovciak2004} and \texttt{relline} \citep{Dauser2010} (setting the black hole spin to zero) for a range of parameters: excellent agreement was found in all cases.

To investigate the features of the occulted line model and the range of parameters over which they give rise to sizeable effects in the profile, we worked out a number of examples.
We adopted a limb-brightened surface emissivity and a radial power law index of $q=-3$, comparing the disc line profile occulted by stars of different radii $\ge 6~r_{g}$ with the unocculted case for which we set $R_{NS} = 3~r_{g}$ (note that any star with $ R_{NS} \lesssim 5~r_{g} $ would produce no occultation features even for a 90-degree inclination). 

\begin{figure}[t]
	\centering
	\hspace{-1.1cm}
	\includegraphics[width=0.75\linewidth]{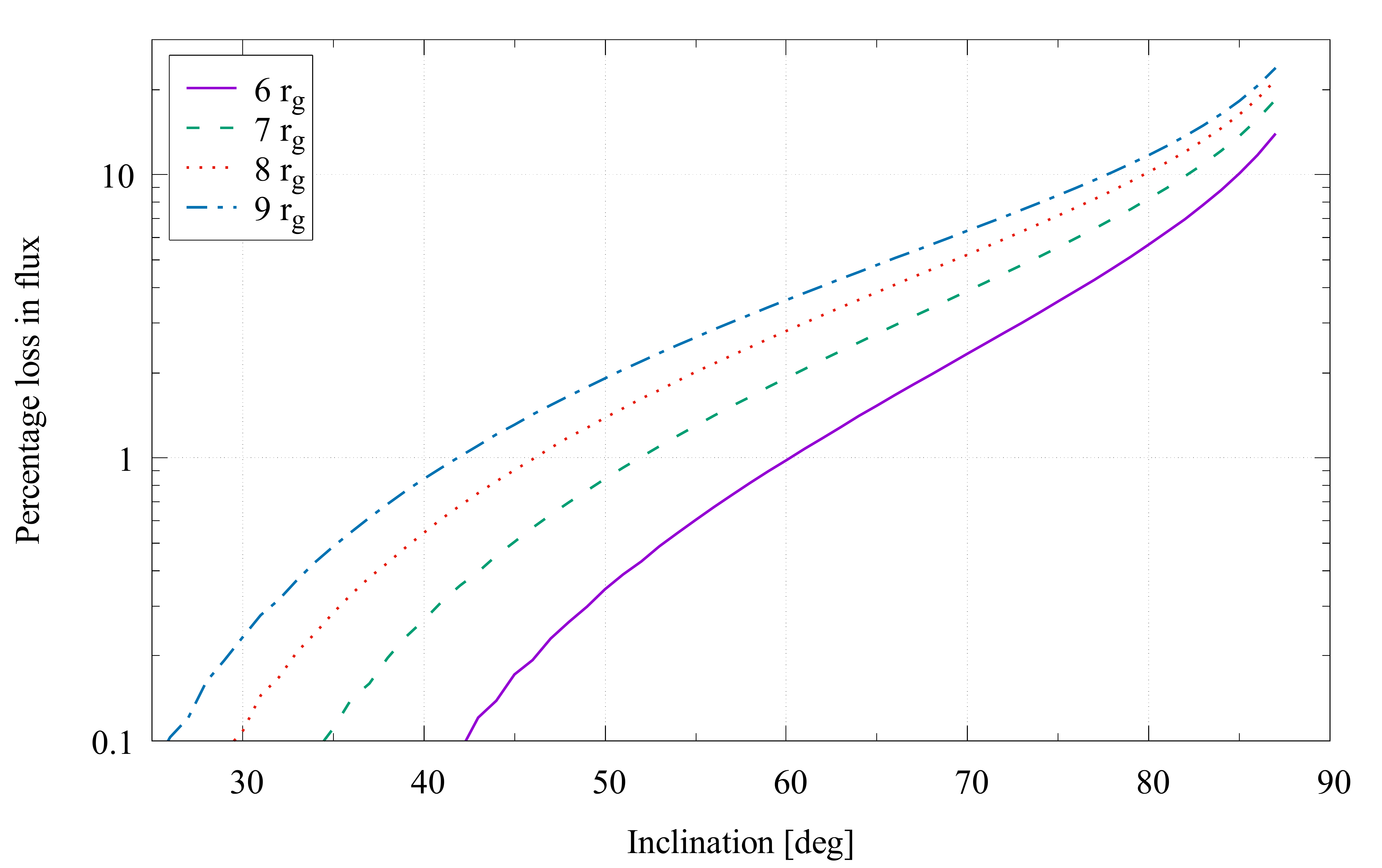}
	\caption{The percentage loss in total flux with respect to the unocculted case is plotted here for different star radii: we used $ q=-3 $ and the limb-brightening law, and the disc radii were set to $ r_{in} = R_{NS}  $, $ r_{out} = 5\cdot10^{5} ~r_{g} $.}
	\label{fig:lostflux}
\end{figure}

The simplest consequence of the occultation of the inner part of the disc is a reduction in the observed flux. The degree to which occultation reduces the line flux relative to the unocculted case is plotted in Fig. \ref{fig:lostflux}: values as high as 10\% can be found for all star radii above $6~r_{g}$ when the disc reaches the star surface, and in all these cases a 2\% reduction in flux is reached for $i<70$\textdegree.
Due to their higher relative flux in the central part of the disc, the isotropic and limb-darkening laws lead to enhanced emission from the far side of the inner disc regions, and thus to pronounced occultation features, compared with the limb-brightening case.

\begin{figure}[t]
	\centering
	\vspace{0.9cm}
	\includegraphics[width=1.0\linewidth]{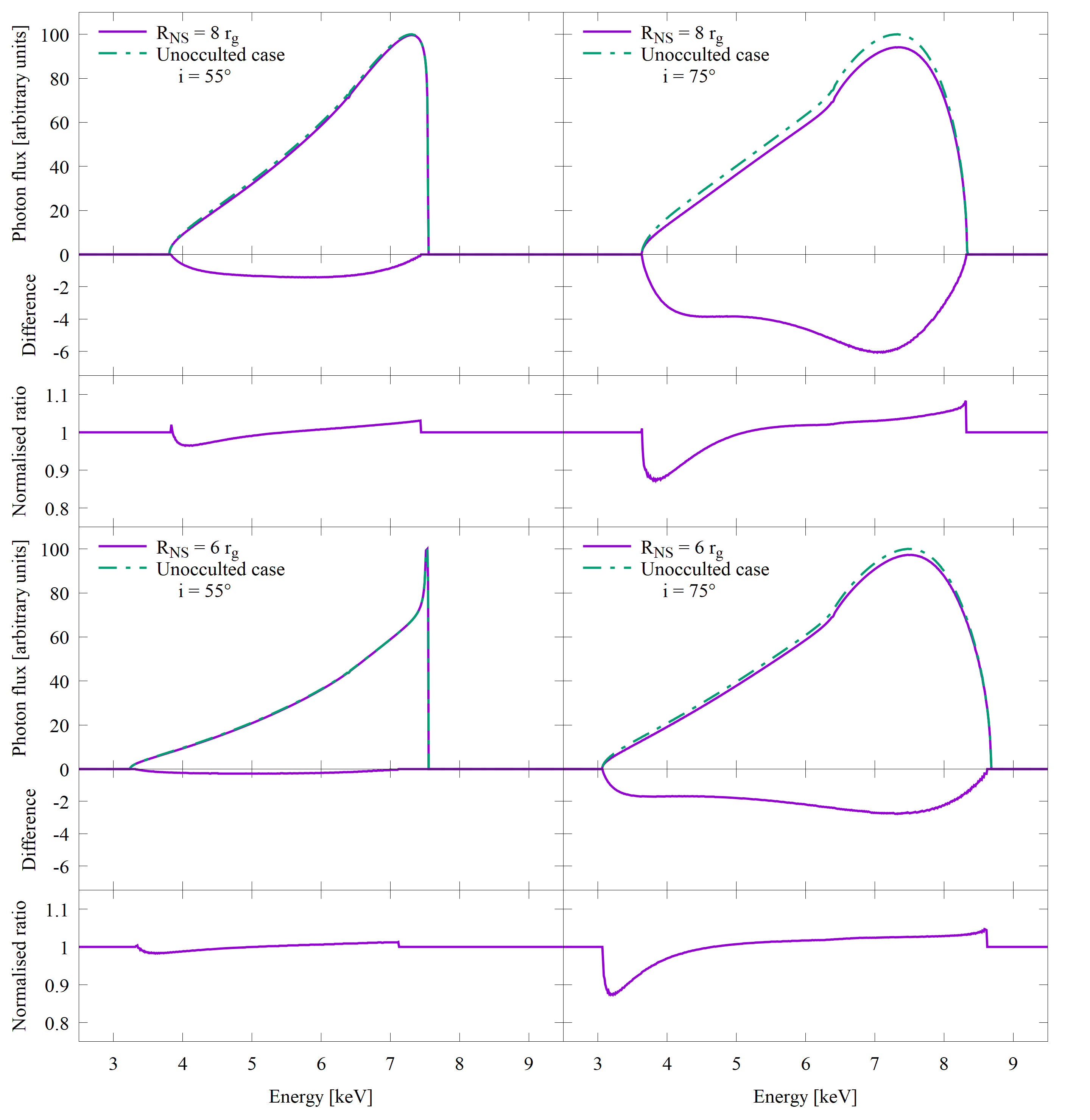}
	\vspace{0.1cm}
	\caption{Occulted and unocculted line profiles for systems observed under an inclination of 55\textdegree\ and 75\textdegree\ (columns), hosting a central star of radius equal to 6 and $ 8~r_{g}$ (rows). In all cases, $r_{in} $ is equal to the occulting star's radius, while $ q=-3 $, $ r_{out} = 5\cdot10^{6} ~r_{g} $ and the limb-brightening angular emission law was used. In each of the four subfigures the three panels show (top to bottom): \textit{i.} the occulted (solid) and unocculted (dot-dashed) line profiles, with the flux expressed in units of the unocculted peak flux; \textit{ii.} the difference between the two profiles, in the same units as before; \textit{iii.} the ratio between the two profiles at each energy bin, normalised to its average value over the whole profile.}
	\label{fig:profiles}
\end{figure}

A mere line flux reduction, not accompanied by a profile change, would result simply in a lower equivalent width and thus remain undetectable in applications to real data. However, clear signatures emerge from a comparison between pairs of occulted and non-occulted line profiles. Four examples are shown in Fig. \ref{fig:profiles} together with their difference and normalised ratio.

The shape of the normalised ratio displays a trough and peak close to the low and high energy end of the profile, respectively: such asymmetry is due to the redshift distribution over the ``crescent'' of the far side of the disc that is occulted by the star. Since in systems observed under low inclinations or containing a small central star the most highly blueshifted photons are not occulted by the star surface, we see that in those profiles the highest energies are virtually unaffected (see, e.g., the bottom-left subfigure in Fig. \ref{fig:profiles}, which represents a system with $R_{NS} = 6~r_{g}$ seen under $i = 55$\textdegree).

Occultation features scale, through Eq.~\ref{eq:almax}, as $R_{NS}/r_{g} \propto R_{NS}/M$: spectral analysis of occulted line profiles provides thus a new method to measure neutron star compactness. 

\section{Data analysis and simulations} \label{sec:datasim}

Through the insertion of \sd into \xspec as a local model \citep{arnaud2018x}, we used the partial occultation model as a tool to constrain or measure the $ R_{NS}/r_{g} $ ratio following two different approaches. In the first we analysed the spectrum from a high-inclination NS LMXB system, \sedi, as observed by the Nuclear Spectroscopic Telescope Array \citep[\textit{NuSTAR},][]{NustarWP}; in the second approach we simulated and analysed X-ray spectra that could be obtained for the same system by next-generation instrumentation such as the Large Area Detector (LAD) on board the Enhanced X-ray Timing and Polarimetry mission (\textit{eXTP}), currently under development and expected to launch in the mid-2020s \citep{Zhang2019}.

\subsection{NuSTAR analysis of \sedi}

\sedi\ is a widely studied LMXB that has been observed in both the soft and hard state \citep[][and references therein]{Giacconi1974, Lyu2014}: its X-ray emission has a flux usually larger than 120 mCrab and shows a broad Fe \kalp \ line, consistent with that produced by a disc extending close to the innermost stable circular orbit (ISCO) at $6~r_g$ \citep[e.g.][]{Ludlam2017}.

The system displays thermonuclear (Type I) X-ray bursts during which a nearly coherent signal at 581~Hz arising from the neutron star rotation is frequently observed \citep{Strohm1998}. Its inclination is generally believed to be higher than 65\textdegree: certain estimates, including some based on the width of the Fe \kalp\ line profile itself, return values of $i \gtrsim 80$\textdegree\ \citep[see, e.g.,][]{Pandel2008,Cackett2010,Lyu2014}. However, this contrasts with the absence of dips or eclipses in the light curve from \sedi, which even allowing for favourable conditions in the system geometry seems to rule out inclination values higher than 75\textdegree\ \citep{Sanna2013}.

\sidecaptionvpos{figure}{c}
\begin{SCfigure}[][b]
	\centering
	\includegraphics[width=0.58\linewidth]{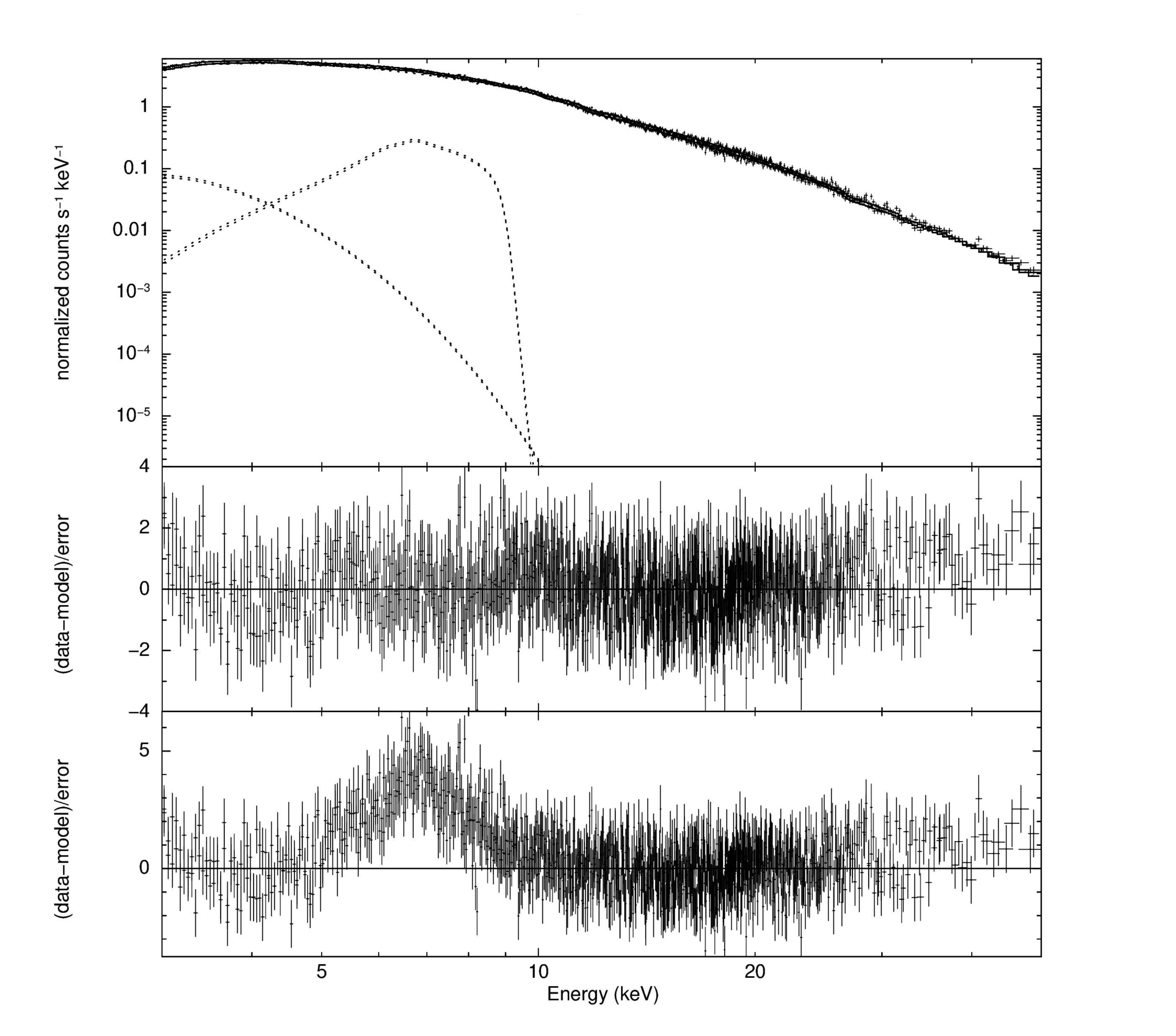}
	\caption{Best-fit \textit{NuSTAR} spectrum of \sedi\ obtained with the combination of \texttt{tbabs}, \texttt{kerrconv}, \texttt{diskbb}, \texttt{pexriv} and \texttt{shaddisk}, and its residuals in the middle panel; the bottom panel shows the residuals when eliminating the iron line, i.e., accounting for the continuum alone.}
\label{fig:ourfit}
\end{SCfigure}

\citet{Ludlam2017} studied the X-ray spectrum of \sedi\ from a 19.8~ks \textit{NuSTAR} observation, looking in particular at the inner disc radius in order to set an upper limit on the NS radius-to-mass ratio. The instrumentation on board the \textit{NuSTAR} mission boasts a 400~eV spectral resolution under 50~keV and is not affected by photon pile-up; it operates in the $ 3\div79$~keV band, with a peak in effective area of 800~cm$^2$ around 10~keV \citep{NustarWP}. When taking the NS to be non-rotating, \citeauthor{Ludlam2017} found best-fit values of $i = 76.5\div79.9$\textdegree \ and $r_{in} = 6.00\div6.36~r_{g}$, which suggests that occultation features in the iron line might be present, although small: therefore we elected to analyse the same observation (ID 30101024002, 06/06/2015) within \textsc{Xspec}.

\sidecaptionvpos{table}{c}
\begin{SCtable}
	\renewcommand{\arraystretch}{0.96}
	\centering
	\hspace{-1.7cm}
	\begin{tabular}[c]{L L L C L}
		\hline
		$ Component $  & $ Parameter $ &  $ Units $  & $Value $\\
		\hline 
		TBabs       & n_{H}     &10^{22}~cm^{-2}&  0.4       & f  \\
		kerrconv    & Index1    &           &  2.45          & l  \\
		kerrconv    & Index2    &           &  2.45          & l  \\
		kerrconv    & r_{br}    &   [r_g]   &  150           & f  \\
		kerrconv    & a         &           &  0.0           & f  \\
		kerrconv    & Incl      &    deg    &  80.3          & l  \\         
		kerrconv    & Rin       & [r_{ISCO}]&  1.0           & l  \\    
		kerrconv    & Rout      & [r_{ISCO}]&  170           & f  \\   
		pexriv      & PhoIndex  &           &  1.772_{-0.013}^{+0.004}          &    \\   
		pexriv      & foldE     &   keV     &  19.26_{-0.15}^{+0.14}          &    \\          
		pexriv      & rel_{refl}&           &  0.34_{-0.05}^{+0.05}          &    \\    
		pexriv      & Redshift  &           &  0.0           & f  \\   
		pexriv      & abund     &           &  2.8           & f  \\ 
		pexriv      & Fe~abund  &           &  2.8           & f  \\ 
		pexriv      & cosIncl   &           &  0.169         & l  \\  
		pexriv      & T_{disk}  &  K        &  1\cdot10^6    & f  \\          
		pexriv      & \xi       &  erg*cm/s &  2.9_{-1.9}^{+2.1}\cdot10^3  &    \\          
		pexriv      & norm      &           &  0.404_{-0.001}^{+0.001}          &    \\ 
		diskbb      & Tin       &   keV     &  0.53_{-0.03}^{+0.02}          &    \\          
		diskbb      & norm      &           &  61.67         & l  \\  
		shaddisk    & LineE     &   keV     &  6.46_{-0.22}^{+0.06}          &    \\         
		shaddisk    & EmissInd  &           &  -2.45_{-0.05}^{+0.05}          &    \\ 
		shaddisk    & Rin       &   [r_g]   &  6.00_{-0.00}^{+0.09}          &    \\ 
		shaddisk    & Rout      &   [r_g]   &  990           & f  \\  
		shaddisk    & Incl      &   deg     &  80.3_{-0.6}^{+1.0}          &    \\         
		shaddisk    & Rns       &   [r_g]   &  5.99_{-5.99}^{+0.09}          &    \\
		shaddisk    & AngDep    &           &  -1.0 $ (limb br.)$& f  \\
		shaddisk    & norm      &           &  2.29_{-0.01}^{+0.01}\cdot10^{-3} &  \\
		\hline 
	\end{tabular}
		\caption{ \\ Best-fit values using \texttt{tbabs}, \texttt{kerrconv}, \texttt{diskbb}, \texttt{pexriv} and \sd on the \textit{NuSTAR} spectrum from \sedi. Parameter values indicated with $ f $ were fixed while the ones indicated with $ l $ were linked to other parameters. $ n_{H} $ is the equivalent column density of hydrogen on the line of sight used to calculate absorption; in \texttt{kerrconv} the two radial emissivity indices are both linked to \texttt{shaddisk}'s, so the break radius after which the routine uses the second one is fixed and irrelevant in this analysis, while the dimensionless spin $ a $ is fixed to 0 and the other parameters are linked to the relevant ones in \texttt{shaddisk}. The first two parameters in \texttt{pexriv} are the illuminating power law photon index and cut-off energy, followed by $ rel_{refl} $ that defines what fraction of its emission is due to the reflected component, the system's redshift and the metal and iron abundances (in units of the solar ones); the cosine of the inclination is linked to $ i $ in \sd while $ T_{disk} $ and $ \xi $ are respectively the disc's temperature and ionisation parameter. Lastly, \texttt{diskbb} only has the disc's temperature at the inner radius as a free parameter, since the normalisation depends on the system's distance, inclination and inner disc radius.}
\label{tab:ffit}
\end{SCtable}

In fitting the spectrum of \sedi\ we added the partially-occulted line profile model (\texttt{shaddisk}) to a relatively simple choice of components consisting of: \texttt{diskbb} for the blackbody emission from the disc \citep[see, e.g.,][]{Mitsuda1984}; \texttt{pexriv} for the exponentially cut off power law spectrum reflected from the ionized material in and above the disc \citep{Magdziarz1995}; \texttt{kerrconv} for the relativistic smearing of the continuum spectrum \citep{Brenneman2006}; \texttt{tbabs} for the absorption along the line of sight \citep{Wilms2000}. 

Our fit results are summarised in Table \ref{tab:ffit}: the reduced $ \chi^2 $ is equal to $\chi^2/$\#dof = 1659/1490 = 1.113, but the shape of the residuals suggests some missing contribution (see Fig.~\hyperref[fig:ourfit]{6}). Let us note that the best-fit values for our main parameters of interest ($ r_{in} $, $ i $ and \rns) are consistent at 90\% confidence with the ones found in \citet{Ludlam2017}, since we find $ r_{in} = 6.01_{-0.01}^{+0.09} ~r_{g}$, $ i = 80.3_{-0.6}^{+1.0}$ degrees and $R_{NS} = 5.99_{-5.99}^{+0.07} ~r_{g}$, which clearly is just an upper limit on the neutron star radius.

The parameters derived from this analysis of \sedi\ suggest that the signal to noise ratio provided by the current X-ray instrumentation is probably insufficient to measure the \rns\ parameter with \texttt{shaddisk} in other high inclination neutron star LMXB: we take, however, the continuum model components and values from Table \ref{tab:ffit} as the basis for our simulated eXTP observations of \sedi.

An extensive application of the new technique to archival X-ray spectra will be presented elsewhere. 
We did not attempt a similar analysis of \sedi\ with the instrumentation on board \textit{XMM-Newton} owing to the problems linked with it, such as dead time and photon pile-up.

\begin{figure}[t]
	\centering
	\gridline{\hspace{-0.2cm}
		\fig{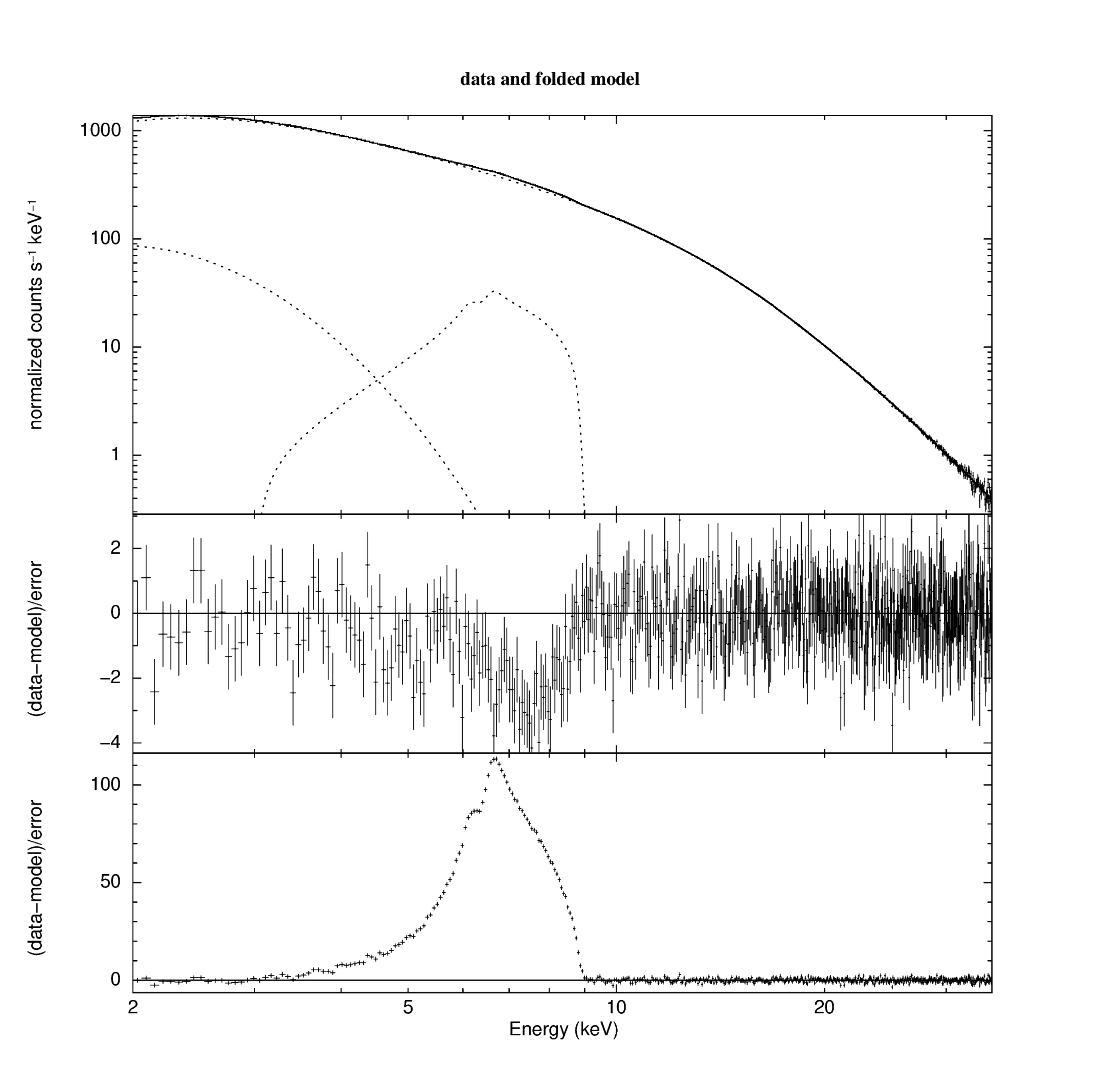}{0.51\textwidth}{}
		\fig{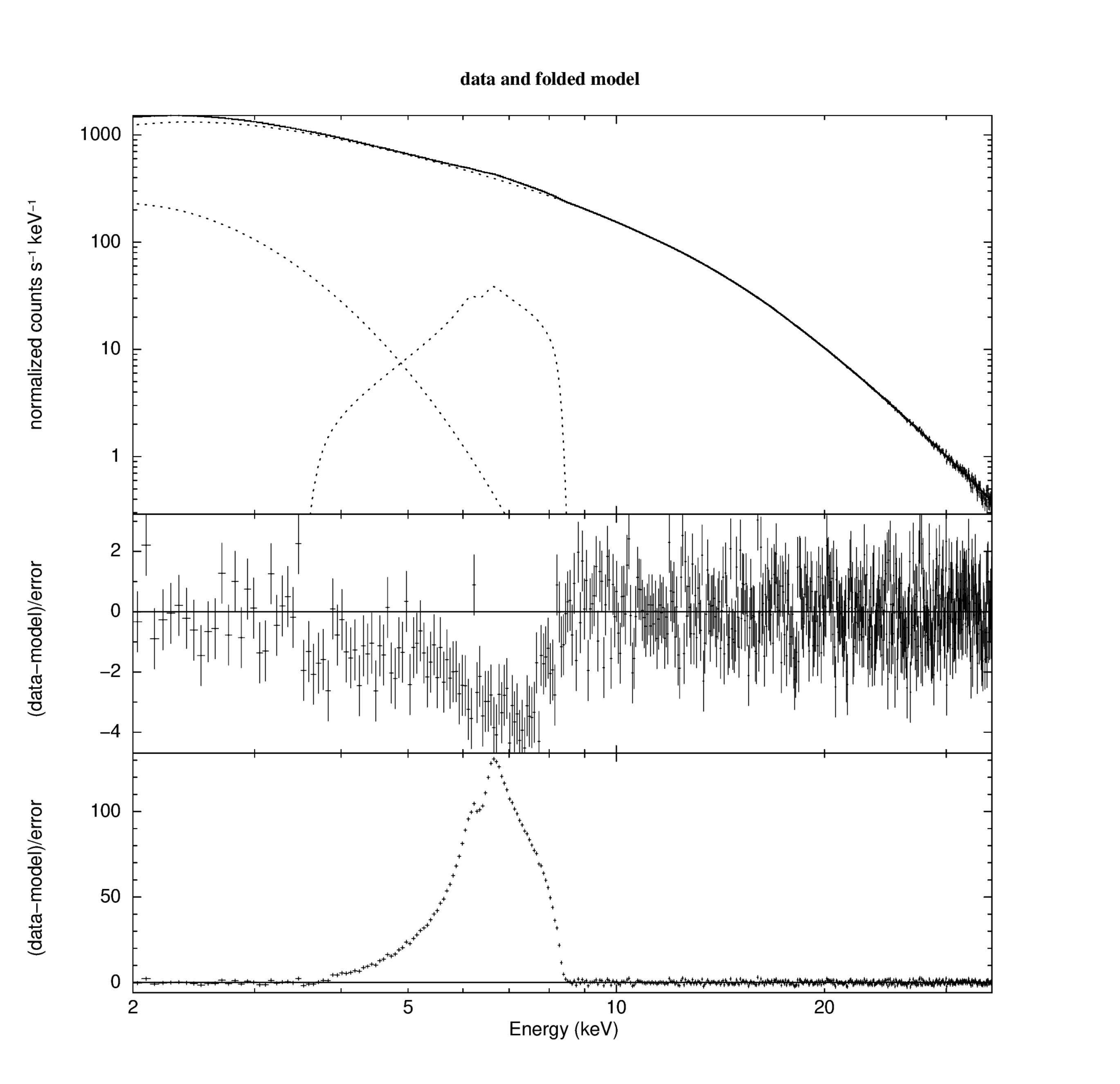}{0.51\textwidth}{}
	} \vspace{-0.5cm}
	\caption{Simulated eXTP spectra from a 100~ks observation of a \textasciitilde120~mCrab system with (left) $ R_{NS} = 6~r_g$, $ r_{in} = 6~r_g$, and $ i = 80$\textdegree\ and (right) $ R_{NS} = 8~r_g$, $ r_{in} = 8~r_g$, $ i = 75$\textdegree: in the two middle panels \rns\  was set to $ 4~r_g$, so the residuals correspond to the deviation produced by not accounting for the occultation; the bottom ones show the residuals with respect to the continuum alone, so that the missing iron line stands out clearly.}
	\label{fig:simres}
\end{figure}

\subsection{Detecting the occultation with future missions}

The \textit{enhanced X-ray Timing and Polarimetry Mission} (eXTP) is a Chinese-led mission being studied for a perspective launch in the mid-2020s \citep{Zhang2019}. Its main instrument, the Large Area Detector (LAD) affords an effective area of $ 3.4$~m$^2$ (about one order of magnitude larger than X-ray instruments of the past and present generations) in the \textasciitilde$8\div10$~keV range and $ 1.5$~m$^2$ at 2 keV, for a total range of $ 2\div 80$ keV; it exploits large monolithic silicon drift detectors (SDDs) with a spectral resolution of \textasciitilde$200$~eV at 6 keV.

Through the use of its public response matrices, we simulated eXTP/LAD observations of \sedi\ within \textsc{Xspec}: for each simulation the integration time was set to 100 ks, while the energy range is constrained on the low end by the capabilities of eXTP: for the aims of this work we limited our spectra to the $ 2\div 35 $ keV range. 

We generated spectra for three different star radii (6, 7 and 8 $ r_g $), setting the inner disc radius first to the surface of the star and then one gravitational radius larger. The inclination was set to five different values for all discs that reached the star surface (60, 65, 70, 75 and 80\textdegree), then to 55\textdegree\ for the $ R_{NS} = 8~r_{g} $ case; when the disc is at $1~r_{g}$ from the surface, we considered at least three possible inclinations for each star radius, with more cases for larger stars. Thus, we produced and fitted a total of 28 different spectra, checking whether \xspec could recover the initial parameter values of $ r_{in} $ and \rns\ within a 90\% confidence region. 

The results are very promising: the contribution given by the occultation can be clearly singled out, as we can see from the residuals in the two middle panels in Fig. \ref{fig:simres} (compare with the middle panels in Fig. \ref{fig:profiles}).

Thanks to the high signal-to-noise ratio of large-area observations, the value of \rns\ can be determined to precision within a few percent: \hyperref[tab:bfpar]{the following tables} (\ref{tab:bfpar}, \ref{tab:bfpar2}) summarise the best-fit values of the \rns\ and $ r_{in} $ parameters for the various simulated spectra, together with their 90\% confidence region.
The actual star radius falls inside this 90\% confidence range for every simulated spectrum except for the \rns$~=7~r_{g}$, $ r_{in} = 7~r_g $, $ i=75 $\textdegree\ case, whose $ 7~r_{g} $ value is slightly higher than the $ 6.96~r_{g} $ upper extremum of the confidence range.

\begin{table}[t]
	\caption{\\Best-fit $ r_{in} $ and \rns \ parameters for the simulations in which $ r_{in} = R_{NS} $.}
	\label{tab:bfpar}
	\renewcommand{\arraystretch}{1.5}
	\centering
	\begin{tabular}{C|CC|CC|CC}
		\toprule
		             & \multicolumn{2}{C|}{R_{NS} = 6~r_{g}}                           & \multicolumn{2}{C|}{R_{NS} = 7~r_{g}}                           & \multicolumn{2}{C}{R_{NS} = 8~r_{g}}                            \\
		{i}             & r_{in}^{bf} & R_{NS}^{bf} & r_{in}^{bf} & R_{NS}^{bf} & r_{in}^{bf} & R_{NS}^{bf} \\ \hline
		55 $\textdegree$  & -                          & -                       & -                          & -                       & 8.00_{-0.02}^{+0.01} $ $   & 7.68_{-0.21}^{+0.32}      \\
		60 $\textdegree$  & 6.00_{-6.00}^{+0.01} $ $   & 5.90_{-0.01}^{+0.11}    & 7.01_{-0.02}^{+0.01} $ $   & 6.85_{-0.19}^{+0.16}    & 8.00_{-0.02}^{+0.01} $ $   & 8.00_{-0.28}^{+0.01}      \\
		65 $\textdegree$  & 6.01_{-0.01}^{+0.01} $ $   & 6.00_{-0.23}^{+0.01}    & 7.00_{-0.01}^{+0.01} $ $   & 7.00_{-0.15}^{+0.01}    & 8.01_{-0.01}^{+0.02} $ $   & 7.99_{-0.30}^{+0.03}      \\
		70 $\textdegree$  & 6.00_{-0.00}^{+0.02} $ $   & 5.96_{-0.22}^{+0.04}    & 7.00_{-0.01}^{+0.01} $ $   & 7.00_{-0.08}^{+0.02}    & 8.00_{-0.00}^{+0.02} $ $   & 7.88_{-0.19}^{+0.17}      \\
		75 $\textdegree$  & 6.00_{-6.00}^{+0.03} $ $   & 5.99_{-0.13}^{+0.17}    & 7.01_{-0.03}^{+0.01} $ $   & 6.86_{-0.15}^{+0.10}    & 8.00_{-0.04}^{+0.01} $ $   & 8.00_{-0.10}^{+0.01}      \\
		80 $\textdegree$  & 6.00_{-6.00}^{+0.01} $ $   & 5.94_{-0.07}^{+0.06}    & 6.99_{-0.02}^{+0.03} $ $   & 6.98_{-0.16}^{+0.16}    & 8.00_{-0.00}^{+0.03} $ $   & 7.99_{-0.08}^{+0.02}      \\ \bottomrule
	\end{tabular}
\vspace{0.5cm}
	\caption{\\Best-fit $ r_{in} $ and \rns \ parameters for the simulations in which $ r_{in} = R_{NS} + 1~r_{g}$.}
	\label{tab:bfpar2}
	\renewcommand{\arraystretch}{1.5}
	\centering
	\begin{tabular}{C|CC|CC|CC}
		\toprule
		& \multicolumn{2}{C|}{R_{NS} = 6~r_{g}}                           & \multicolumn{2}{C|}{R_{NS} = 7~r_{g}}                           & \multicolumn{2}{C}{R_{NS} = 8~r_{g}}                            \\
		i  & r_{in}^{bf} & R_{NS}^{bf} & r_{in}^{bf} & R_{NS}^{bf} & r_{in}^{bf} & R_{NS}^{bf} \\ \hline
		60 $\textdegree$ & -                          & -                       & -                          & -                       & 9.00_{-0.03}^{+0.02}  $ $  & 8.13_{-0.34}^{+0.23}    \\
		65 $\textdegree$ & -                          & -                       & 8.00_{-0.03}^{+0.02} $ $   & 7.07_{-7.07}^{+0.31}    & 8.95_{-0.01}^{+0.01}  $ $  & 8.15_{-0.40}^{+0.24}    \\
		70 $\textdegree$ & 7.00_{-0.01}^{+0.01} $ $   & 5.99_{-5.99}^{+0.82}    & 8.00_{-0.01}^{+0.02} $ $   & 7.16_{-0.20}^{+0.18}    & 8.97_{-0.01}^{+0.03}  $ $  & 7.82_{-0.23}^{+0.27}    \\
		75 $\textdegree$ & 6.99_{-0.01}^{+0.02} $ $   & 5.99_{-5.99}^{+0.36}    & 8.05_{-0.03}^{+0.03} $ $   & 7.12_{-0.22}^{+0.22}    & 8.91_{-0.04}^{+0.04}  $ $  & 7.87_{-0.32}^{+0.28}    \\
		80 $\textdegree$ & 7.00_{-0.01}^{+0.01} $ $   & 6.05_{-0.15}^{+0.20}    & 7.97_{-0.01}^{+0.01} $ $   & 7.06_{-0.16}^{+0.14}    & 8.95_{-0.02}^{+0.02}  $ $  & 8.07_{-0.18}^{+0.18}    \\ \bottomrule
	\end{tabular}
\end{table}

Some clarifications are in order; first of all the presence of some values whose lower limit is zero: in the case of the disc inner radius in systems with a $ 6~r_{g} $ star, this is due to $ 6~r_{g} $ already being the hard limit for the parameter, which makes it impossible for \xspec to find a lower value. For the simulations with \rns$~=6~r_{g}$, $ r_{in} = 7~r_g $ and \rns$~=7~r_{g}$, $ r_{in} = 8~r_g $ instead, the difficulty in finding a lower limit on \rns\ is given by the rather small (or absent) effect caused by the occultation: larger stars allow the right value of $R_{NS}$ to be determined, as we can see from the $R_{NS} = 8~r_{g}$, $ r_{in} = 9~r_{g} $ simulations. 

In all except the aforementioned cases with $ r_{in} = R_{NS} + 1~r_{g}$, 90\%-confidence errors on the neutron star radius are always under 5\% of the \rns\ value in either direction, the average being 2.2\%; furthermore, most of them are below 3.5\%. 
This is the precision expected to be necessary to put tight constraints on the equation of state of supranuclear density matter based on neutron star parameters \citep{Watts2019}.

Moreover, we find that in all the simulations in which \xspec only finds an upper limit on the star radius, this constraint in more stringent than the one imposed by the inner disc radius alone: this is true of both the cases with a $7~r_{g}$ and a $6~r_{g}$ star whose accretion disc does not touch the surface and it is another confirmation that the occultation effects (or in this case the lack thereof) can provide a strong limit on the radius-to-mass ratio.
These results make us optimistic about the possibilities of future observations by large-area telescopes of both new and already observed systems to better constrain the \rns$/r_{g}$ ratio and therefore the equation of state of neutron stars, hoping to shed more light on the behaviour of ultradense matter. 

\section{Discussion and future perspectives}

In this work we introduced a new technique to measure the radius-to-mass ratio of disc-accreting non-magnetic neutron stars in X-ray binary systems by exploiting the features imprinted by occultation by the body of the star itself on the profile of the relativistically broadened and redshifted Fe \kalp\ originating from the disc.
We developed a fast \xspec routine (\texttt{shaddisk}) for the integration of the line profile that can be used efficiently to fit the X-ray spectra observed from these neutron stars. 

We investigated the key characteristics of the technique, determining the conditions under which the trajectory of disc Fe \kalp\ line photons intercepts the neutron star's surface, 
and calculated the Fe \kalp\ line profiles for range of NS radii and disc inclinations, by using different prescriptions for the disc angular emissivity law. These profiles were compared with the corresponding unocculted ones, which allowed us to determine that occultation alters significantly the line profile for neutron star radii $ \gtrsim 6\div7~GM/c^2 \simeq 9\div10~(M/M_{\odot}) $ km and disc inclinations $ \gtrsim 65\div70 $\textdegree. 

The analysis of the \textit{NuSTAR} X-ray spectrum of the LMXB system \sedi\ by using \sd did not provide statistically significant evidence that its Fe \kalp\ line profile is affected by occultation by the neutron star body. Only an upper limit on the star radius could be found, consistent with that derived from the inner disc radius (\textasciitilde$6~r_g $).

In order to check whether the occultation features can be detected in the Fe \kalp\ line by very large effective area X-ray instruments of the next generation, we carried out extensive simulations by using the response matrices of the Large Area Detector, to be flown on board eXTP. The spectral parameters of \sedi\ were used in these simulations, except that the line profile was varied over a range of disc inclinations and neutron star radii.  
The fit to the simulated spectra with \sd demonstrated that the occultation features can be revealed in the Fe \kalp\ profiles over a wide range of parameters, resulting in $ 2\div 3 \%$ precise measurements of the NS radius-to-mass ratio in most cases ($\leq5$\% in all cases).

\begin{figure}[t!]
	\centering
	\makebox[\textwidth][c]{%
		\centering
		\includegraphics[width=0.80\linewidth]{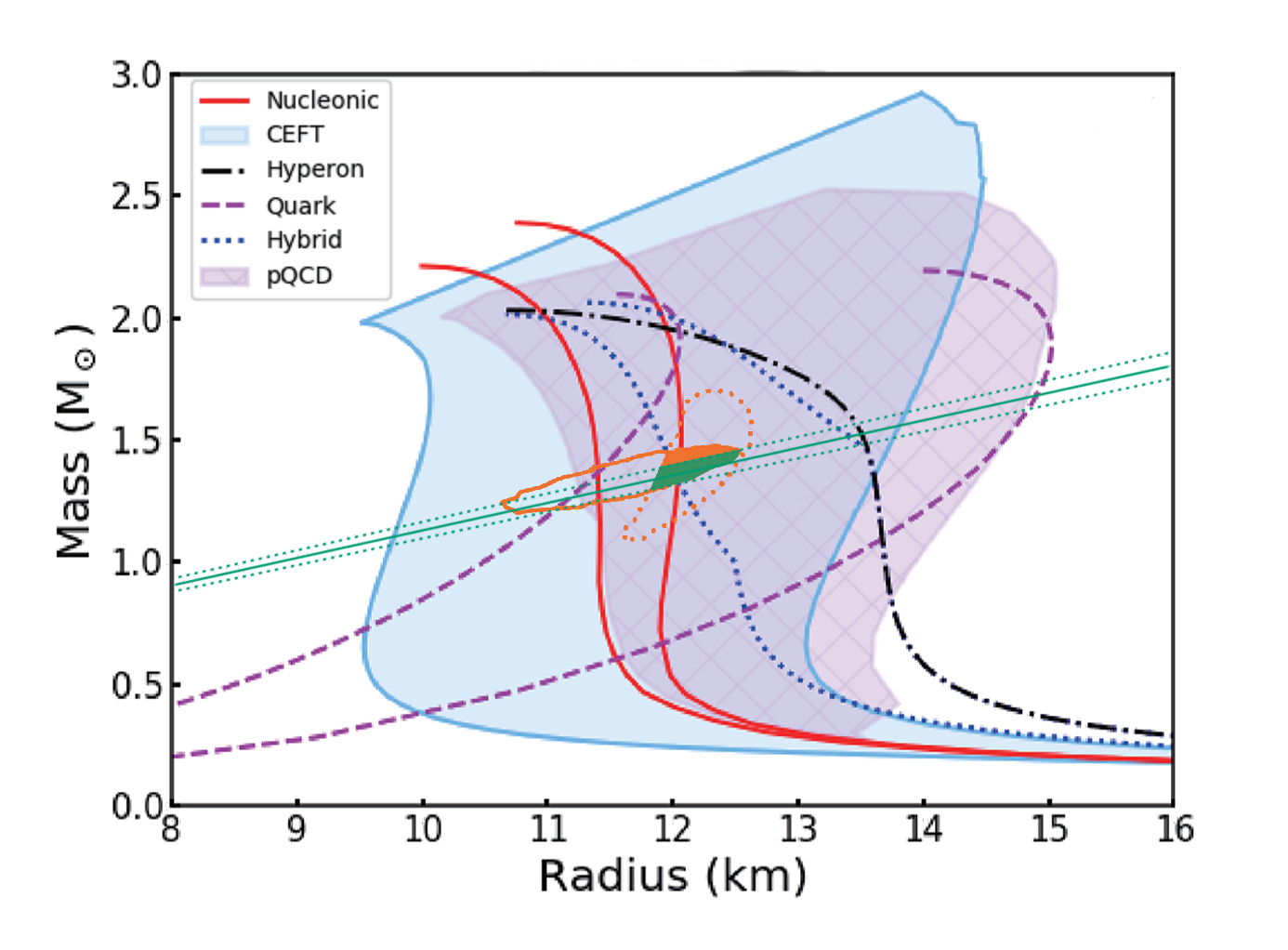}
	}%
	\caption{Limits on the mass-radius plane obtained with \sd if an $ R_{NS}/r_{g} $ ratio of $ 6\pm0.18 $ is found (green lines): the errors correspond to 3\% precision and are similar to the ones of the \rns$~=6~r_{g} = r_{in}$, $ i=75 $\textdegree\ simulation in Table \ref{tab:bfpar}. The underlying graph corresponds to a composition of Fig. 3 and 7 in \citet{Watts2019}, in which the open curves and areas indicated in the legend represent various equations of state, with different colours and styles for different underlying microphysics \citep[for their description, see][]{Watts2019}. The orange solid curve (shifted down \textasciitilde$0.1~M_{\odot}$ for a clearer comparison) represents the kind of constraint we can obtain with eXTP from the modelling of the pulse profiles from localised hotspots on the star: when intersected with the confidence region that could be found through the analysis of atmosphere spectral models \citep[orange dotted, shifted down too][]{Nattila2017}, this gives the filled area in orange. The simultaneous use of the three methods provides an even tighter constraint on the NS size (filled region in green).}
	\label{fig:concl2}
\end{figure}

Some of the assumptions and limitations of our model are to be addressed: the occultation model that we developed implicitly assumes that the neutron star is non-rotating and spherical, and the innermost accretion disc region has negligible thickness.
According to standard theory, the thickness-to-radius ratio $h/R$ of the inner, radiation-pressure dominated region of the disc is of order $h/R \propto L/L_{Edd} $~\textasciitilde~$ 1\% $ for X-ray sources, like \sedi, whose luminosity is about 1\% of the Eddington limit \citep{ShakuraSunyaev1973}. Therefore finite disc thickness will alter only to a very small extent the geometry of the occulted inner region of disc.

Rotation induces oblateness, such that the occulting surface of the neutron star is that of an oblate spheroid, no longer that of a sphere. For the high end of the spin frequency distribution of accreting NSs \citep[\textasciitilde600 Hz, as in the case of \sedi,][]{Papitto2014,Patruno2017} the equatorial radius is about $ 5\div15 \% $ larger than the polar radius, depending on the star's mass and EoS \citep[see, e.g.,][]{Friedman1986,Morsink2007}. The neutron star's oblateness will thus alter the outer border of the occulted inner disc ``crescent'' by making it slightly thinner in the middle and more elongated and thicker at the extremes: the Fe-line profiles will be slightly altered correspondingly.
The rotation of the star (and the resulting oblateness) also alters the spacetime close to it with frame dragging effects as well as terms arising from higher order mass multipoles: these in turn will affect matter and photon geodesics.  
However, these two effects have opposing contributions on both the position of the ISCO and the motion of the innermost disc regions \citep{Morsink1999, Bhattacharyya2011}: the appendix provides approximate estimates of their impact, showing that for highly spinning NSs the analysis of partially occulted lines returns the value of the equatorial radius-to-mass ratio, which ensures that the technique would retain nearly the same level of accuracy even when applied to more realistic systems.

We conclude that the new method provides an innovative, independent way of measuring the radius-to-mass ratio of neutron stars with next generation large area X-ray instruments. Though limited to large enough neutron stars in binary systems seen from relatively high inclinations, the method will afford the few percent precision that is required to gather quantitative information on the EoS of matter at supranuclear densities \citep{Watts2019}; its precision is also comparable to that of other methods that will be exploited in the same time frame by using X-ray instrumentation of the same type, such as the eXTP/LAD.  
This is illustrated in Fig. \ref{fig:concl2}, where the constraint from our method intersect that from two other techniques which exploit eXTP's instrumentation. It is apparent that the size of the allowed region would reduce considerably. 

We note that the model and code we developed can be employed also to search for Fe \kalp\ partial occultation features originating from other bodies than the neutron star surface. For instance a spherical, optically thick corona such as that envisaged as the primary continuum X-ray source in models of disc-accreting compact objects would imprint features in the iron line that can be successfully singled out by the new technique.
A detailed study of such systems will be presented in future works.

Moreover, the development a convolution model based on \sd for the relativistic smearing and partial occultation of different emission features in LMXB spectra is currently being finalised. 

\acknowledgments

ADR, LS and TDS acknowledge financial contribution from the agreement ASI-INAF n.2017-14-H.O. 
LS also acknowledges financial contributions from the ASI-INAF agreement I/037/12/0 and from the ``iPeska'' research grant (P.I. Andrea Possenti) funded under the INAF call PRIN-SKA/CTA (resolution 70/2016).
AP acknowledges financial support from grants ASI/INAF I/037/12/0, ASI/INAF 2017-14-H.0 (PI: Belloni) and from INAF ``Sostegno alla ricerca scientifica main streams dell'INAF'', Presidential Decree 43/2018 (PI: Belloni).

\software{\xspec \citep{Arnaud1996}, HEAsoft \citep{Heasoft2014}, relline \citep{Dauser2010}, kyrline \citep{Dovciak2004}, rns \citep{Stergioulas1995}, LSDplus \citep{Bakala2015}}

\

\appendix \label{appendix}

\section*{Estimate of smaller effects on the occulted line profiles}

We investigate here the effects of the star's oblateness, rotation and of its mass multipoles by comparing the line profiles obtained with \sd in the Schwarzschild metric to the ones produced by an advanced numerical code, \textit{Lensing Simulation Device plus} \citep[LSDplus,][]{Bakala2015}, starting by the simple case of an oblate star in the Schwarzschild metric. For a given equatorial radius, an oblate star occults less than a spherical one, and as expected we are able to observe more of the flux around the centre of the broadened line profile, since matter in the area behind the NS moves almost perpendicularly to the line of sight. In particular, we considered an ellipticity $ e = \sqrt{1-R^2_{pol}/R^2_{eq}} $ of $e=0.3$, corresponding to the rotational frequency $f=600$ Hz using the empirical formula \citep[see Eq. 8. in][]{Morsink2007} based on the modelling of NS structure through the {\bf rns} code \citep{Stergioulas1995}: here, $ R_{pol} $ and $ R_{eq}$ are, respectively, the polar and equatorial radii of the star. Choosing the same X-ray continuum we used for the eXTP simulations, we generated the line profile occulted by the oblate star under LSDplus; the corresponding X-ray spectrum simulated with \xspec features a disc going from the surface of the star ($R_{eq} = 6~r_g$) to $ 60~r_g $, observed under an inclination $ i = 70$\textdegree\ for 100 ks by eXTP. Fitting this profile with \texttt{shaddisk} (which assumes the star to be spherical) we obtain a best-fit star radius of $ 5.70^{+0.12}_{-0.11}~r_g $, which makes this case consistent with a slightly smaller star: in particular, since the polar radius of the oblate star is equal to \textasciitilde$5.72~r_g$, this best-fit value suggests that the occultation features on the line profile are more sensitive to the polar radius than to the equatorial one in the Schwarzschild metric.

For a second, more complete comparison, we needed to account for the influence the star rotation and oblateness on the motion of disc matter.
We adopted the formula for the Keplerian angular velocity on circular orbits in Hartle-Thorne geometry describing the external spacetime in the vicinity of a rotating neutron star. Recall that the Hartle-Thorne metric is an vacuum exact solution of the Einstein field equations that describes the exterior of any slowly and rigidly rotating, stationary and axially symmetric body. The metric is given with accuracy up to the second order terms in the body's dimensionless angular momentum (spin) $j = J/M^2$, and first order in terms of its dimensionless quadrupole momentum, $q =-Q/M^3$ \citep[see, e.g.,][]{HartleThorne1968,Abramowicz2003,Urbancova2019}. The Keplerian angular velocity in the Hartle-Thorne geometry on the corotating circular geodesics with radius $r$ is
\begin{equation}
\Omega_{HT}=\Omega_{K}\,\left(1 - j\,(r_{g}/r)^{3/2}+j^2\,F_1(r)+q\,F_2(r)\right)\,
\end{equation}
where $\Omega_{K}$ is the Keplerian angular velocity in the Schwarzschild geometry and the coefficients $F_1(r)$, $F_2(r)$ are given by Eqs. (17-20) in \citet{Abramowicz2003}.
The Hartle-Thorne angular velocity $\Omega_{HT}$ increases with growing oblateness (i.e. increasing dimensionless quadrupole momentum $q$), but decreases with increasing spin $j$ \citep[for further details, see][]{Urbancova2019}.

The Kerr metric in the Boyer-Lindquist coordinates describing stable rotating black hole corresponds the Hartle-Thorne metric after putting $q = j^2$ and applying a subtle coordinate transformation described by Eqs. (11) and (12) in \citet{Abramowicz2003}. However, in the case of a neutron star the quadrupole momentum value is strongly related to the EOS of ultradense matter. Numerical modelling of NS structure show that the value of dimensionless quadrupole momentum $q$ can vary in the $(2\div 12) j^2$ range and increases with growing stiffness of EOS \citep{Arnett1977,Laarakkers1999,Morsink1999}. For a NS with mass $M=1.4 M_\odot$, rotational frequency $f=600$ Hz and spin $j=0.3$, we chose $q=0.7$, close to the maximum dimensionless quadrupole moment \citep[see Table IV in][]{Laarakkers1999}. As with the previous case, we approximate the oblate shape of the NS with a spheroid characterised by an eccentricity $e=0.3$ and $ R_{eq} = 6~r_g $. 

By generating another profile with LSDplus for this configuration and adding it to the usual continuum, we simulated another eXTP observation that we could fit with \texttt{shaddisk}. The resulting best-fit value of $R_{NS} = 6.00^{+0.03}_{-0.12}~r_g $ is consistent with the value of the equatorial radius owing to the interplay among frame-dragging effects, oblateness and the additional contribution of the mass quadrupole \cite[see also][for similar results on the inner disc radius around rotating NSs]{Bhattacharyya2011}. This consistency, although it introduces a degeneracy in discriminating between the rotating and non-rotating cases, ensures that the accuracy of this new method is essentially maintained when considering all the relevant effects. 

\bibliography{newbib}

\begin{thebibliography}{}
\expandafter\ifx\csname natexlab\endcsname\relax\def\natexlab#1{#1}\fi
\providecommand{\url}[1]{\href{#1}{#1}}
\providecommand{\dodoi}[1]{doi:~\href{http://doi.org/#1}{\nolinkurl{#1}}}
\providecommand{\doeprint}[1]{\href{http://ascl.net/#1}{\nolinkurl{http://ascl.net/#1}}}
\providecommand{\doarXiv}[1]{\href{https://arxiv.org/abs/#1}{\nolinkurl{https://arxiv.org/abs/#1}}}

\bibitem[{{Abbott} {et~al.}(2018){Abbott}, {Abbott}, {Abbott}, {Acernese},
  {Ackley}, {Adams}, {Adams}, {Addesso}, {Adhikari}, \&
  {Adya}}]{Abbott2018Mass}
{Abbott}, B.~P., {Abbott}, R., {Abbott}, T.~D., {et~al.} 2018, \prl, 121,
  161101, \dodoi{10.1103/PhysRevLett.121.161101}

\bibitem[{{Abramowicz} {et~al.}(2003){Abramowicz}, {Almergren}, {Kluzniak}, \&
  {Thampan}}]{Abramowicz2003}
{Abramowicz}, M.~A., {Almergren}, G.~J.~E., {Kluzniak}, W., \& {Thampan}, A.~V.
  2003, arXiv e-prints, gr.
\newblock \doarXiv{gr-qc/0312070}

\bibitem[{{Akmal} {et~al.}(1998){Akmal}, {Pandharipande}, \&
  {Ravenhall}}]{Akmal1998}
{Akmal}, A., {Pandharipande}, V.~R., \& {Ravenhall}, D.~G. 1998, \prc, 58,
  1804, \dodoi{10.1103/PhysRevC.58.1804}

\bibitem[{Arnaud {et~al.}(2018)Arnaud, Gordon, \& Dorman}]{arnaud2018x}
Arnaud, K., Gordon, C., \& Dorman, B. 2018, An X-Ray Spectral Fitting Package

\bibitem[{{Arnaud}(1996)}]{Arnaud1996}
{Arnaud}, K.~A. 1996, in Astronomical Society of the Pacific Conference Series,
  Vol. 101, Astronomical Data Analysis Software and Systems V, ed. G.~H.
  {Jacoby} \& J.~{Barnes}, 17

\bibitem[{{Arnett} \& {Bowers}(1977)}]{Arnett1977}
{Arnett}, W.~D., \& {Bowers}, R.~L. 1977, \apjs, 33, 415,
  \dodoi{10.1086/190434}

\bibitem[{{Bakala} {et~al.}(2015){Bakala}, {Goluchov{\'a}}, {T{\"o}r{\"o}k},
  {{\v{S}}r{\'a}mkov{\'a}}, {Abramowicz}, {Vincent}, \& {Mazur}}]{Bakala2015}
{Bakala}, P., {Goluchov{\'a}}, K., {T{\"o}r{\"o}k}, G., {et~al.} 2015, \aap,
  581, A35, \dodoi{10.1051/0004-6361/201525867}

\bibitem[{Bao {et~al.}(1994)Bao, Hadrava, \& Ostgaard}]{Bao1994}
Bao, G., Hadrava, P., \& Ostgaard, E. 1994, The Astrophysical Journal, 435, 55,
  \dodoi{10.1086/174793}

\bibitem[{Beloborodov(2002)}]{Beloborodov2002}
Beloborodov, A.~M. 2002, \apjl, 566, L85, \dodoi{10.1086/339511}

\bibitem[{{Bhattacharyya}(2010)}]{Bhattacharyya2010}
{Bhattacharyya}, S. 2010, Advances in Space Research, 45, 949,
  \dodoi{10.1016/j.asr.2010.01.010}

\bibitem[{{Bhattacharyya}(2011)}]{Bhattacharyya2011}
---. 2011, \mnras, 415, 3247, \dodoi{10.1111/j.1365-2966.2011.18936.x}

\bibitem[{{Brenneman} \& {Reynolds}(2006)}]{Brenneman2006}
{Brenneman}, L.~W., \& {Reynolds}, C.~S. 2006, \apj, 652, 1028,
  \dodoi{10.1086/508146}

\bibitem[{{Burgay} {et~al.}(2003){Burgay}, {D'Amico}, {Possenti}, {Manchester},
  {Lyne}, {Joshi}, {McLaughlin}, {Kramer}, {Sarkissian}, \&
  {Camilo}}]{Burgay2003}
{Burgay}, M., {D'Amico}, N., {Possenti}, A., {et~al.} 2003, \nat, 426, 531,
  \dodoi{10.1038/nature02124}

\bibitem[{{Cackett} {et~al.}(2010){Cackett}, {Miller}, {Ballantyne}, {Barret},
  {Bhattacharyya}, {Boutelier}, {Miller}, {Strohmayer}, \&
  {Wijnands}}]{Cackett2010}
{Cackett}, E.~M., {Miller}, J.~M., {Ballantyne}, D.~R., {et~al.} 2010, \apj,
  720, 205, \dodoi{10.1088/0004-637X/720/1/205}

\bibitem[{{Chandrasekhar}(1950)}]{Chandrasekhar1950}
{Chandrasekhar}, S. 1950, {Radiative transfer} (Clarendon Press)

\bibitem[{Chandrasekhar(1983)}]{Chandrasekhar1983}
Chandrasekhar, S. 1983, The mathematical theory of black holes (Clarendon
  Press/Oxford University Press).
\newblock \url{http://adsabs.harvard.edu/abs/1983mtbh.book.....C}

\bibitem[{{Chatziioannou} {et~al.}(2017){Chatziioannou}, {Clark}, {Bauswein},
  {Millhouse}, {Littenberg}, \& {Cornish}}]{Chatziioannou2017}
{Chatziioannou}, K., {Clark}, J.~A., {Bauswein}, A., {et~al.} 2017, \prd, 96,
  124035, \dodoi{10.1103/PhysRevD.96.124035}

\bibitem[{{Chen} \& {Eardley}(1991)}]{ChenEardley1991}
{Chen}, K., \& {Eardley}, D.~M. 1991, \apj, 382, 125, \dodoi{10.1086/170701}

\bibitem[{{Cottam} {et~al.}(2002){Cottam}, {Paerels}, \& {Mendez}}]{Cottam2002}
{Cottam}, J., {Paerels}, F., \& {Mendez}, M. 2002, \nat, 420, 51,
  \dodoi{10.1038/nature01159}

\bibitem[{{Dauser} {et~al.}(2010){Dauser}, {Wilms}, {Reynolds}, \&
  {Brenneman}}]{Dauser2010}
{Dauser}, T., {Wilms}, J., {Reynolds}, C.~S., \& {Brenneman}, L.~W. 2010,
  \mnras, 409, 1534, \dodoi{10.1111/j.1365-2966.2010.17393.x}

\bibitem[{{De Falco} {et~al.}(2016){De Falco}, Falanga, \& Stella}]{DFS2016}
{De Falco}, V., Falanga, M., \& Stella, L. 2016, Astronomy {\&} Astrophysics,
  595, A38, \dodoi{10.1051/0004-6361/201629075}

\bibitem[{{Dov{\v c}iak} {et~al.}(2004){Dov{\v c}iak}, {Karas}, \&
  {Yaqoob}}]{Dovciak2004}
{Dov{\v c}iak}, M., {Karas}, V., \& {Yaqoob}, T. 2004, \apjs, 153, 205,
  \dodoi{10.1086/421115}

\bibitem[{Fabian {et~al.}(1989)Fabian, Rees, Stella, \& White}]{Fabian1989}
Fabian, A.~C., Rees, M.~J., Stella, L., \& White, N.~E. 1989, Monthly Notices
  of the Royal Astronomical Society, 238, 729, \dodoi{10.1093/mnras/238.3.729}

\bibitem[{{Friedman} {et~al.}(1986){Friedman}, {Ipser}, \&
  {Parker}}]{Friedman1986}
{Friedman}, J.~L., {Ipser}, J.~R., \& {Parker}, L. 1986, \apj, 304, 115,
  \dodoi{10.1086/164149}

\bibitem[{{Fukue} \& {Akizuki}(2006)}]{Fukue2006}
{Fukue}, J., \& {Akizuki}, C. 2006, \pasj, 58, 1039,
  \dodoi{10.1093/pasj/58.6.1039}

\bibitem[{{George} \& {Fabian}(1991)}]{George1991}
{George}, I.~M., \& {Fabian}, A.~C. 1991, \mnras, 249, 352,
  \dodoi{10.1093/mnras/249.2.352}

\bibitem[{{Ghisellini} {et~al.}(1994){Ghisellini}, {Haardt}, \&
  {Matt}}]{Ghisellini1994}
{Ghisellini}, G., {Haardt}, F., \& {Matt}, G. 1994, \mnras, 267, 743,
  \dodoi{10.1093/mnras/267.3.743}

\bibitem[{{Giacconi} {et~al.}(1974){Giacconi}, {Murray}, {Gursky}, {Kellogg},
  {Schreier}, {Matilsky}, {Koch}, \& {Tananbaum}}]{Giacconi1974}
{Giacconi}, R., {Murray}, S., {Gursky}, H., {et~al.} 1974, The Astrophysical
  Journal Supplement Series, 27, 37, \dodoi{10.1086/190288}

\bibitem[{{Glendenning} \& {Schaffner-Bielich}(1999)}]{Glendenning1999}
{Glendenning}, N.~K., \& {Schaffner-Bielich}, J. 1999, \prc, 60, 025803,
  \dodoi{10.1103/PhysRevC.60.025803}

\bibitem[{{Goosmann} {et~al.}(2007){Goosmann}, {Mouchet}, {Czerny}, {Dov{\v
  c}iak}, {Karas}, {R{\'o}{\v z}a{\'n}ska}, \& {Dumont}}]{Goosmann2007}
{Goosmann}, R.~W., {Mouchet}, M., {Czerny}, B., {et~al.} 2007, \aap, 475, 155,
  \dodoi{10.1051/0004-6361:20078273}

\bibitem[{{Haardt}(1993)}]{Haardt1993}
{Haardt}, F. 1993, \apj, 413, 680, \dodoi{10.1086/173036}

\bibitem[{{Haensel} {et~al.}(2009){Haensel}, {Zdunik}, {Bejger}, \&
  {Lattimer}}]{Haensel2009}
{Haensel}, P., {Zdunik}, J.~L., {Bejger}, M., \& {Lattimer}, J.~M. 2009, \aap,
  502, 605, \dodoi{10.1051/0004-6361/200811605}

\bibitem[{{Harrison} {et~al.}(2013){Harrison}, {Craig}, {Christensen},
  {Hailey}, {Zhang}, {Boggs}, {Stern}, {Cook}, {Forster}, {Giommi},
  {Grefenstette}, {Kim}, {Kitaguchi}, {Koglin}, {Madsen}, {Mao}, {Miyasaka},
  {Mori}, {Perri}, {Pivovaroff}, {Puccetti}, {Rana}, {Westergaard}, {Willis},
  {Zoglauer}, {An}, {Bachetti}, {Barri{\`e}re}, {Bellm}, {Bhalerao},
  {Brejnholt}, {Fuerst}, {Liebe}, {Markwardt}, {Nynka}, {Vogel}, {Walton},
  {Wik}, {Alexander}, {Cominsky}, {Hornschemeier}, {Hornstrup}, {Kaspi},
  {Madejski}, {Matt}, {Molendi}, {Smith}, {Tomsick}, {Ajello}, {Ballantyne},
  {Balokovi{\'c}}, {Barret}, {Bauer}, {Blandford}, {Brandt}, {Brenneman},
  {Chiang}, {Chakrabarty}, {Chenevez}, {Comastri}, {Dufour}, {Elvis}, {Fabian},
  {Farrah}, {Fryer}, {Gotthelf}, {Grindlay}, {Helfand}, {Krivonos}, {Meier},
  {Miller}, {Natalucci}, {Ogle}, {Ofek}, {Ptak}, {Reynolds}, {Rigby},
  {Tagliaferri}, {Thorsett}, {Treister}, \& {Urry}}]{NustarWP}
{Harrison}, F.~A., {Craig}, W.~W., {Christensen}, F.~E., {et~al.} 2013, \apj,
  770, 103, \dodoi{10.1088/0004-637X/770/2/103}

\bibitem[{{Hartle} \& {Thorne}(1968)}]{HartleThorne1968}
{Hartle}, J.~B., \& {Thorne}, K.~S. 1968, \apj, 153, 807,
  \dodoi{10.1086/149707}

\bibitem[{Heasarc(2014)}]{Heasoft2014}
Heasarc, N. H. E. A. S. A. R.~C. 2014, {HEAsoft: Unified Release of FTOOLS and
  XANADU}.
\newblock \doeprint{1408.004}

\bibitem[{{Heinke} {et~al.}(2014){Heinke}, {Cohn}, {Lugger}, {Webb}, {Ho},
  {Anderson}, {Campana}, {Bogdanov}, {Haggard}, {Cool}, \&
  {Grindlay}}]{Heinke2014}
{Heinke}, C.~O., {Cohn}, H.~N., {Lugger}, P.~M., {et~al.} 2014, \mnras, 444,
  443, \dodoi{10.1093/mnras/stu1449}

\bibitem[{{Hinderer} {et~al.}(2010){Hinderer}, {Lackey}, {Lang}, \&
  {Read}}]{Hinderer2010}
{Hinderer}, T., {Lackey}, B.~D., {Lang}, R.~N., \& {Read}, J.~S. 2010, \prd,
  81, 123016, \dodoi{10.1103/PhysRevD.81.123016}

\bibitem[{{Israel} {et~al.}(2005){Israel}, {Belloni}, {Stella}, {Rephaeli},
  {Gruber}, {Casella}, {Dall'Osso}, {Rea}, {Persic}, \&
  {Rothschild}}]{Israel2005}
{Israel}, G.~L., {Belloni}, T., {Stella}, L., {et~al.} 2005, \apjl, 628, L53,
  \dodoi{10.1086/432615}

\bibitem[{{Kehl} {et~al.}(2016){Kehl}, {Wex}, {Kramer}, \& {Liu}}]{Kehl2016}
{Kehl}, M.~S., {Wex}, N., {Kramer}, M., \& {Liu}, K. 2016, arXiv e-prints,
  arXiv:1605.00408.
\newblock \doarXiv{1605.00408}

\bibitem[{{La Placa} {et~al.}(2019){La Placa}, Bakala, Stella, \&
  Falanga}]{ResNote}
{La Placa}, R., Bakala, P., Stella, L., \& Falanga, M. 2019, Research Notes of
  the {AAS}, 3, 99, \dodoi{10.3847/2515-5172/ab3227}

\bibitem[{{Laarakkers} \& {Poisson}(1999)}]{Laarakkers1999}
{Laarakkers}, W.~G., \& {Poisson}, E. 1999, \apj, 512, 282,
  \dodoi{10.1086/306732}

\bibitem[{{Laor}(1991)}]{Laor1991}
{Laor}, A. 1991, \apj, 376, 90, \dodoi{10.1086/170257}

\bibitem[{{Lattimer} \& {Prakash}(2004)}]{LattPrak2004}
{Lattimer}, J.~M., \& {Prakash}, M. 2004, Science, 304, 536,
  \dodoi{10.1126/science.1090720}

\bibitem[{{Lattimer} \& {Prakash}(2016)}]{LattPrak2016}
---. 2016, \physrep, 621, 127, \dodoi{10.1016/j.physrep.2015.12.005}

\bibitem[{{Lattimer} \& {Schutz}(2005)}]{LattSchutz2005}
{Lattimer}, J.~M., \& {Schutz}, B.~F. 2005, \apj, 629, 979,
  \dodoi{10.1086/431543}

\bibitem[{{Lewin} {et~al.}(1993){Lewin}, {van Paradijs}, \& {Taam}}]{Lewin1993}
{Lewin}, W.~H.~G., {van Paradijs}, J., \& {Taam}, R.~E. 1993, \ssr, 62, 223,
  \dodoi{10.1007/BF00196124}

\bibitem[{{Lindblom}(1992)}]{Lindblom1992}
{Lindblom}, L. 1992, \apj, 398, 569, \dodoi{10.1086/171882}

\bibitem[{{Ludlam} {et~al.}(2017){Ludlam}, {Miller}, {Bachetti}, {Barret},
  {Bostrom}, {Cackett}, {Degenaar}, {Di Salvo}, {Natalucci}, {Tomsick},
  {Paerels}, \& {Parker}}]{Ludlam2017}
{Ludlam}, R.~M., {Miller}, J.~M., {Bachetti}, M., {et~al.} 2017, \apj, 836,
  140, \dodoi{10.3847/1538-4357/836/1/140}

\bibitem[{{Lyne} {et~al.}(2004){Lyne}, {Burgay}, {Kramer}, {Possenti},
  {Manchester}, {Camilo}, {McLaughlin}, {Lorimer}, {D'Amico}, \&
  {Joshi}}]{Lyne2004}
{Lyne}, A.~G., {Burgay}, M., {Kramer}, M., {et~al.} 2004, Science, 303, 1153,
  \dodoi{10.1126/science.1094645}

\bibitem[{{Lyu} {et~al.}(2014){Lyu}, {M{\'e}ndez}, {Sanna}, {Homan}, {Belloni},
  \& {Hiemstra}}]{Lyu2014}
{Lyu}, M., {M{\'e}ndez}, M., {Sanna}, A., {et~al.} 2014, \mnras, 440, 1165,
  \dodoi{10.1093/mnras/stu279}

\bibitem[{{Magdziarz} \& {Zdziarski}(1995)}]{Magdziarz1995}
{Magdziarz}, P., \& {Zdziarski}, A.~A. 1995, \mnras, 273, 837,
  \dodoi{10.1093/mnras/273.3.837}

\bibitem[{{Martocchia} \& {Matt}(1996)}]{Martocchia1996}
{Martocchia}, A., \& {Matt}, G. 1996, \mnras, 282, L53,
  \dodoi{10.1093/mnras/282.4.L53}

\bibitem[{{Matt} {et~al.}(1991){Matt}, {Perola}, \& {Piro}}]{MattPP1991}
{Matt}, G., {Perola}, G.~C., \& {Piro}, L. 1991, \aap, 247, 25

\bibitem[{Misner {et~al.}(1973)Misner, Thorne, \& Wheeler}]{Misner1973}
Misner, C.~W., Thorne, K.~S., \& Wheeler, J.~A. 1973, Gravitation (San
  Francisco: W.H.~Freeman and Co., 1973).
\newblock \url{http://adsabs.harvard.edu/abs/1973grav.book.....M}

\bibitem[{{Mitsuda} {et~al.}(1984){Mitsuda}, {Inoue}, {Koyama}, {Makishima},
  {Matsuoka}, {Ogawara}, {Shibazaki}, {Suzuki}, {Tanaka}, \&
  {Hirano}}]{Mitsuda1984}
{Mitsuda}, K., {Inoue}, H., {Koyama}, K., {et~al.} 1984, \pasj, 36, 741

\bibitem[{{Morsink} {et~al.}(2007){Morsink}, {Leahy}, {Cadeau}, \&
  {Braga}}]{Morsink2007}
{Morsink}, S.~M., {Leahy}, D.~A., {Cadeau}, C., \& {Braga}, J. 2007, \apj, 663,
  1244, \dodoi{10.1086/518648}

\bibitem[{{Morsink} \& {Stella}(1999)}]{Morsink1999}
{Morsink}, S.~M., \& {Stella}, L. 1999, \apj, 513, 827, \dodoi{10.1086/306876}

\bibitem[{{M{\"u}ller} \& {Serot}(1996)}]{Muller1996}
{M{\"u}ller}, H., \& {Serot}, B.~D. 1996, \nphysa, 606, 508,
  \dodoi{10.1016/0375-9474(96)00187-X}

\bibitem[{{M{\"u}ther} {et~al.}(1987){M{\"u}ther}, {Prakash}, \&
  {Ainsworth}}]{Muther1987}
{M{\"u}ther}, H., {Prakash}, M., \& {Ainsworth}, T.~L. 1987, Physics Letters B,
  199, 469, \dodoi{10.1016/0370-2693(87)91611-X}

\bibitem[{{N{\"a}ttil{\"a}} {et~al.}(2017){N{\"a}ttil{\"a}}, {Miller},
  {Steiner}, {Kajava}, {Suleimanov}, \& {Poutanen}}]{Nattila2017}
{N{\"a}ttil{\"a}}, J., {Miller}, M.~C., {Steiner}, A.~W., {et~al.} 2017, \aap,
  608, A31, \dodoi{10.1051/0004-6361/201731082}

\bibitem[{{N{\"a}ttil{\"a}} \& {Pihajoki}(2018)}]{Nattila2018}
{N{\"a}ttil{\"a}}, J., \& {Pihajoki}, P. 2018, \aap, 615, A50,
  \dodoi{10.1051/0004-6361/201630261}

\bibitem[{{{\"O}zel} \& {Freire}(2016)}]{Ozel2016}
{{\"O}zel}, F., \& {Freire}, P. 2016, \araa, 54, 401,
  \dodoi{10.1146/annurev-astro-081915-023322}

\bibitem[{{{\"O}zel} \& {Psaltis}(2009)}]{Ozel2009}
{{\"O}zel}, F., \& {Psaltis}, D. 2009, \prd, 80, 103003,
  \dodoi{10.1103/PhysRevD.80.103003}

\bibitem[{{Pandel} {et~al.}(2008){Pandel}, {Kaaret}, \& {Corbel}}]{Pandel2008}
{Pandel}, D., {Kaaret}, P., \& {Corbel}, S. 2008, \apj, 688, 1288,
  \dodoi{10.1086/592429}

\bibitem[{{Papitto} {et~al.}(2014){Papitto}, {Torres}, {Rea}, \&
  {Tauris}}]{Papitto2014}
{Papitto}, A., {Torres}, D.~F., {Rea}, N., \& {Tauris}, T.~M. 2014, \aap, 566,
  A64, \dodoi{10.1051/0004-6361/201321724}

\bibitem[{{Patruno} {et~al.}(2017){Patruno}, {Haskell}, \&
  {Andersson}}]{Patruno2017}
{Patruno}, A., {Haskell}, B., \& {Andersson}, N. 2017, \apj, 850, 106,
  \dodoi{10.3847/1538-4357/aa927a}

\bibitem[{{Raithel} {et~al.}(2018){Raithel}, {{\"O}zel}, \&
  {Psaltis}}]{Raithel2018}
{Raithel}, C.~A., {{\"O}zel}, F., \& {Psaltis}, D. 2018, \apjl, 857, L23,
  \dodoi{10.3847/2041-8213/aabcbf}

\bibitem[{{Reynolds}(2014)}]{Reynolds2014}
{Reynolds}, C.~S. 2014, \ssr, 183, 277, \dodoi{10.1007/s11214-013-0006-6}

\bibitem[{{Reynolds} \& {Nowak}(2003)}]{Reynolds2003}
{Reynolds}, C.~S., \& {Nowak}, M.~A. 2003, \physrep, 377, 389,
  \dodoi{10.1016/S0370-1573(02)00584-7}

\bibitem[{{Rezzolla} {et~al.}(2018){Rezzolla}, Pizzochero, Jones, Rea, \&
  Vida{\~n}a}]{ASSL2018}
{Rezzolla}, L., Pizzochero, P., Jones, D., Rea, N., \& Vida{\~n}a, I., eds.
  2018, Astrophysics and Space Science Library, Vol. 457, {The Physics and
  Astrophysics of Neutron Stars}

\bibitem[{{Sanna} {et~al.}(2013){Sanna}, {Hiemstra}, {M{\'e}ndez},
  {Altamirano}, {Belloni}, \& {Linares}}]{Sanna2013}
{Sanna}, A., {Hiemstra}, B., {M{\'e}ndez}, M., {et~al.} 2013, \mnras, 432,
  1144, \dodoi{10.1093/mnras/stt530}

\bibitem[{{Shakura} \& {Sunyaev}(1973)}]{ShakuraSunyaev1973}
{Shakura}, N.~I., \& {Sunyaev}, R.~A. 1973, \aap, 24, 337

\bibitem[{{Steiner} \& {Watts}(2009)}]{Steiner2009}
{Steiner}, A.~W., \& {Watts}, A.~L. 2009, \prl, 103, 181101,
  \dodoi{10.1103/PhysRevLett.103.181101}

\bibitem[{{Stergioulas} \& {Friedman}(1995)}]{Stergioulas1995}
{Stergioulas}, N., \& {Friedman}, J.~L. 1995, \apj, 444, 306,
  \dodoi{10.1086/175605}

\bibitem[{{Strohmayer} {et~al.}(1998){Strohmayer}, {Zhang}, {Swank}, {White},
  \& {Lapidus}}]{Strohm1998}
{Strohmayer}, T.~E., {Zhang}, W., {Swank}, J.~H., {White}, N.~E., \& {Lapidus},
  I. 1998, \apj, 498, L135, \dodoi{10.1086/311322}

\bibitem[{{Svoboda}(2010)}]{SvobodaPhD}
{Svoboda}, J. 2010, PhD thesis, Charles University in Prague

\bibitem[{{Urbancov{\'a}} {et~al.}(2019){Urbancov{\'a}}, {Urbanec},
  {T{\"o}r{\"o}k}, {Stuchl{\'\i}k}, {Blaschke}, \& {Miller}}]{Urbancova2019}
{Urbancov{\'a}}, G., {Urbanec}, M., {T{\"o}r{\"o}k}, G., {et~al.} 2019, \apj,
  877, 66, \dodoi{10.3847/1538-4357/ab1b4c}

\bibitem[{{Watts} {et~al.}(2016){Watts}, {Andersson}, {Chakrabarty}, {Feroci},
  {Hebeler}, {Israel}, {Lamb}, {Miller}, {Morsink}, {{\"O}zel}, {Patruno},
  {Poutanen}, {Psaltis}, {Schwenk}, {Steiner}, {Stella}, {Tolos}, \& {van der
  Klis}}]{Watts2016}
{Watts}, A.~L., {Andersson}, N., {Chakrabarty}, D., {et~al.} 2016, Reviews of
  Modern Physics, 88, 021001, \dodoi{10.1103/RevModPhys.88.021001}

\bibitem[{{Watts} {et~al.}(2019){Watts}, {Yu}, {Poutanen}, {Zhang},
  {Bhattacharyya}, {Bogdanov}, {Ji}, {Patruno}, {Riley}, {Bakala}, {Baykal},
  {Bernardini}, {Bombaci}, {Brown}, {Cavecchi}, {Chakrabarty}, {Chenevez},
  {Degenaar}, {Del Santo}, {Di Salvo}, {Doroshenko}, {Falanga}, {Ferdman},
  {Feroci}, {Gambino}, {Ge}, {Greif}, {Guillot}, {Gungor}, {Hartmann},
  {Hebeler}, {Heger}, {Homan}, {Iaria}, {Zand}, {Kargaltsev}, {Kurkela}, {Lai},
  {Li}, {Li}, {Li}, {Linares}, {Lu}, {Mahmoodifar}, {M{\'e}ndez}, {Coleman
  Miller}, {Morsink}, {N{\"a}ttil{\"a}}, {Possenti}, {Prescod- Weinstein},
  {Qu}, {Riggio}, {Salmi}, {Sanna}, {Santangelo}, {Schatz}, {Schwenk}, {Song},
  {{\v{S}}r{\'a}mkov{\'a}}, {Stappers}, {Stiele}, {Strohmayer}, {Tews},
  {Tolos}, {T{\"o}r{\"o}k}, {Tsang}, {Urbanec}, {Vacchi}, {Xu}, {Xu}, {Zane},
  {Zhang}, {Zhang}, {Zhang}, {Zheng}, \& {Zhou}}]{Watts2019}
{Watts}, A.~L., {Yu}, W., {Poutanen}, J., {et~al.} 2019, Science China Physics,
  Mechanics, and Astronomy, 62, 29503, \dodoi{10.1007/s11433-017-9188-4}

\bibitem[{{Wilkins}(2018)}]{Wilkins2018}
{Wilkins}, D.~R. 2018, \mnras, 475, 748, \dodoi{10.1093/mnras/stx3167}

\bibitem[{{Wilms} {et~al.}(2000){Wilms}, {Allen}, \& {McCray}}]{Wilms2000}
{Wilms}, J., {Allen}, A., \& {McCray}, R. 2000, \apj, 542, 914,
  \dodoi{10.1086/317016}

\bibitem[{{Zhang} {et~al.}(2019){Zhang}, {Santangelo}, {Feroci}, {Xu}, {Lu},
  {Chen}, {Feng}, {Zhang}, {Brandt}, {Hernanz}, {Baldini}, {Bozzo}, {Campana},
  {De Rosa}, {Dong}, {Evangelista}, {Karas}, {Meidinger}, {Meuris}, {Nand ra},
  {Pan}, {Pareschi}, {Orleanski}, {Huang}, {Schanne}, {Sironi}, {Spiga},
  {Svoboda}, {Tagliaferri}, {Tenzer}, {Vacchi}, {Zane}, {Walton}, {Wang},
  {Winter}, {Wu}, {in't Zand}, {Ahangarianabhari}, {Ambrosi}, {Ambrosino},
  {Barbera}, {Basso}, {Bayer}, {Bellazzini}, {Bellutti}, {Bertucci},
  {Bertuccio}, {Borghi}, {Cao}, {Cadoux}, {Campana}, {Ceraudo}, {Chen}, {Chen},
  {Chevenez}, {Civitani}, {Cui}, {Cui}, {Dauser}, {Del Monte}, {Di Cosimo},
  {Diebold}, {Doroshenko}, {Dovciak}, {Du}, {Ducci}, {Fan}, {Favre},
  {Fuschino}, {G{\'a}lvez}, {Gao}, {Ge}, {Gevin}, {Grassi}, {Gu}, {Gu}, {Han},
  {Hong}, {Hu}, {Ji}, {Jia}, {Jiang}, {Kennedy}, {Kreykenbohm}, {Kuvvetli},
  {Labanti}, {Latronico}, {Li}, {Li}, {Li}, {Li}, {Li}, {Limousin}, {Liu},
  {Liu}, {Lu}, {Luo}, {Macera}, {Malcovati}, {Martindale}, {Michalska}, {Meng},
  {Minuti}, {Morbidini}, {Muleri}, {Paltani}, {Perinati}, {Picciotto},
  {Piemonte}, {Qu}, {Rachevski}, {Rashevskaya}, {Rodriguez}, {Schanz}, {Shen},
  {Sheng}, {Song}, {Song}, {Sgro}, {Sun}, {Tan}, {Uttley}, {Wang}, {Wang},
  {Wang}, {Wang}, {Wang}, {Wang}, {Watts}, {Wen}, {Wilms}, {Xiong}, {Yang},
  {Yang}, {Yang}, {Yu}, {Zhang}, {Zampa}, {Zampa}, {Zdziarski}, {Zhang},
  {Zhang}, {Zhang}, {Zhang}, {Zhang}, {Zhang}, {Zhang}, {Zhang}, {Zhao},
  {Zheng}, {Zhou}, {Zorzi}, \& {Zwart}}]{Zhang2019}
{Zhang}, S., {Santangelo}, A., {Feroci}, M., {et~al.} 2019, Science China
  Physics, Mechanics, and Astronomy, 62, 29502,
  \dodoi{10.1007/s11433-018-9309-2}

\end{thebibliography}

\end{document}